\documentclass{article}
\usepackage[utf8]{inputenc}
\usepackage{multirow,array}
\usepackage{amsmath, graphicx}

\title{Spatial Games of Fake News}
\author{Matthew I Jones, Scott D Pauls, Feng Fu}

\begin{document}

\maketitle

\section*{Abstract}
To curb the spread of fake news on social media platforms, recent studies have considered an online crowdsourcing fact-checking approach as one possible intervention method to reduce misinformation. However, it remains unclear under what conditions crowdsourcing fact-checking efforts deter the spread of misinformation. To address this issue, we model such distributed fact-checking as `peer policing' that will reduce the perceived payoff to share or disseminate false information (fake news) and also reward the spread of trustworthy information (real news). By simulating our model on synthetic square lattices and small-world networks, we show that the presence of social network structure enables fake news spreaders to be self-organized into echo chambers, thereby providing a boost to the efficacy of fake news and thus its resistance to fact-checking efforts. Additionally, to study our model in a more realistic setting, we utilize a Twitter network dataset and study the effectiveness of deliberately choosing specific individuals to be fact-checkers. We find that targeted fact-checking efforts can be highly effective, seeing the same level of success with as little as a fifth of the number of fact-checkers, but it depends on the structure of the network in question. In the limit of weak selection, we obtain closed-form analytical conditions for critical threshold of crowdsourced fact-checking  in terms of the payoff values in our fact-checker/fake news game.
Our work has practical implications for developing model-based mitigation strategies for controlling the spread of misinformation that interferes with the political discourse.

\section*{Significance Statement}
Modern social media has been inundated by false and misleading headlines and articles. We advance the study of this fake news by developing a game-theoretic model of the spread of fake news in a social network. We are able to test the effectiveness of peer fact-checkers and we estimate that the structure of a social network increases misinformation's resistance to fact-checking in a population by a factor of two to three. Additionally, we find that depending on the structure of the network, carefully choosing which individuals are equipped to be fact-checkers can significantly boost fact-checking efforts.

\section{Background}
In a social media environment that encourages sharing and reposting flashy headlines instead of researching and developing a nuanced understanding of a topic, social media platforms seem to facilitate the spread of fake news \cite{lazer_baum_benkler_berinsky_greenhill_menczer_metzger_nyhan_pennycook_rothschild_schudson_sloman_sunstein_thorson_watts_zittrain_2018,shin_jian_driscoll_bar_2018,vosoughi_roy_aral_2018}. Social influence, following, and unfollowing can create polarized and segregated structure in social media like Twitter~\cite{wang_sirianni_tang_zheng_fu_2020}. These echo chambers create conditions for confirmation bias and selection bias~\cite{poole_rosenthal_1985} and thus can facilitate the spread of misinformation~\cite{delvicario_bessi_zollo_petroni_scala_caldarelli_stanley_quattrociocchi_2016}. Moreover, during the COVID-19 pandemic, misinformation has severely impacted our efforts to control the pandemic (``misinfodemics'') \cite{mian_khan_2020,bursztyn_rao_roth_yanagizawa-drott_2020,pennycook_mcphetres_zhang_lu_rand_2020}. 

In the context of modern politics, partisanship has come to dominate the political sphere and stall political consensus, both amongst the political elite and the general population~\cite{andris_lee_hamilton_martino_gunning_selden_2015,pewreseachcenter_2017b,pewreseachcenter_2014,pewreseachcenter_2016a,pewreseachcenter_2016b}. It has become a major research concern to effectively understand circumstances that will lead to consensus of opinion and others that will lead to divergence of opinion and a weakening of information transfer~\cite{fu_wang_2008,holme_newman_2006,zanette_gil_2006,nardini_kozma_barrat_2008,friedkin_proskurnikov_tempo_parsegov_2016,wang_rong_wu_2016,antonopoulos_shang_2018}. 

In the last decade, the study of misinformation has grown rapidly in an effort to alienate these societal ills. Ref.~\cite{shin_jian_driscoll_bar_2018} traced the lifecycle of 17 popular political rumors that circulated on Twitter over 13 months during the 2012 U.S. presidential election; they found that misinformation tends to come back multiple times after the initial publication, while facts do not. Using massive Twitter datasets, it is recently reported that the spread of true and false news follow distinctive patterns: falsehood diffused significantly faster, deeper, and more broadly than the truth in all categories of information~\cite{vosoughi_roy_aral_2018}. Because almost all news media is advertiser-driven, content publishers are incentivized to spread false information to increase engagement from consumers~\cite{stewart_arechar_rand_plotkin_2021}, so we will focus on fact-checking at the level of the individual consumer.

Ref.~\cite{evans_fu_2018} investigated opinion formation on dynamic social networks through the lens of coevolutionary games~\cite{perc_szolnoki_2010}, and using the voting records of the United States House of Representatives over a timespan of decades, the work presented and validated the conditions for the emergence of partisan echo chambers~\cite{pewreseachcenter_2014,pewreseachcenter_2016a,schmidt_zollo_scala_betsch_quattrociocchi_2018}. Integrating  publicly available Twitter data with an agent-based model of opinion formation driven by socio-cognitive biases, Ref.~\cite{wang_sirianni_tang_zheng_fu_2020} recently found that open-mindedness of individuals is a key determinant of forming echo chambers under dueling campaign influence. 

While it is hard to measure the exact impact of these false stories, it is clear that something about the structure of social media (e.g. reposting/retweeting network) is allowing their spread to continue relatively unchecked. To attempt to quantify the effect network structure has on the proliferation of fake news, we develop a mathematical model of fake news sharing and test it on a variety of social networks. There is an established tradition of using spatial game theory to study problems of coordination and collective action, particularly the Prisoner's Dilemma. The structure of a network has been found to impact the behavior of the system to reinforce good behavior~\cite{ohtsuki_hauert_lieberman_nowak_2006, tarnita_ohtsuki_antal_fu_nowak_2009}, and the evolution of the system can exhibit interesting spatial phenomenon that is not present in the well-mixed case~\cite{nowak_may_1992}. We use a similar strategy to study the spread of fake news through a social network. 

A recent study suggested using an online crowdsourced fact-checking approach as one possible intervention to reduce misinformation \cite{pennycook_rand_2019}. Inspired by this empirical work, here we study spatial games of fake news by modeling distributed fact-checking efforts like `peer policing' which will reduce the perceived payoff to share or disseminate false information (fake news) while rewarding the spread of trustworthy information (real news). Fact-checkers will be placed into the population to model the effect of peer policing efforts. Our agent-based model, where individuals can share real or fake news depending on the behavior and success of their neighbors, is studied with simulations as well as a rigorous mathematical analysis. We find that the presence of subtle network structures known as echo chambers impede crowdsourced fact-checking, thereby requiring a much higher critical distribution threshold of fact-checkers across the population.

\section{Methods \& Model}

The foundation of our model is inspired by the virtual interactions that occur repeatedly on online social media sites. To begin, an individual posts a news story (either true or false) for all of their friends or followers to see. Those who believe the story is true can react positively to the post by liking or sharing, while those who disagree may simply ignore the post or even attempt to debunk a false story by pointing out flaws or sharing a link to a fact-checking website, potentially causing embarrassment and inflicting a penalty on those who share fake news.

A key focus of this chapter is peer fact-checking as opposed to institutional fact-checking. Recent research shows that ``inoculation'' with exposure to a weakened version of misleading arguments (similar to vaccination ideas) is effective at reducing susceptibility to misleading persuasion and thus confers psychological resistance to fake news~\cite{roozenbeek_vanderlinden_2019a, roozenbeek_vanderlinden_2019b, mcguire_papageorgis_1962,banas_rains_2010}. By training some subset of the population to identify and respond to fake news, we can create a decentralized fact-checking system where innoculated individuals will be positioned to apply pressure to their social neighbors that share fake news while also supporting their real news sharing neighbors. Inspired by ``zealot models'' from the field of opinion dynamics \cite{wang_rong_wu_2016}, we assume that these fact-checkers will never change belief because of the behavior of those around them, having been successfully inoculated against fake news,

Consider a network with a two-layer structure: the \emph{spreader layer} describes the information sharing dynamics among spreaders of real news vs. fake news. Unfortunately, fake news tends to spread more effectively than real news on social media~\cite{vosoughi_roy_aral_2018}, so our model gives a higher payoff to individuals sharing fake news in simulations (our analytic results, on the other hand, work for any set of prescribed payoffs). If a spreader of real news (labelled $A$) interacts with another $A$, they both receive a moderate payoff; if a spreader of fake news (labelled $B$) interacts with another $B$, they both receive a slightly larger payoff; if an $A$ interacts with a $B$, both receive a very small (or possibly negative) payoff. To contain the spread of fake news, the natural advantage given to $B$ players will have to be counterbalanced by a penalty inflicted by fact-checkers. The second layer of the social network describes the enforcement infrastructure where these designated fact-checkers, denoted by $C$, perform distributed fact-checking to their spreader neighbors: they provide a reward to $A$ and a harsh penalty to $B$. We assume the proportion of fact-checkers $p_C$ is prescribed and static, representing the level of innoculation effort. The payoff to fact-checkers is irrelevant as the fact-checker population is static, so for simplicity we set it to zero. A selection strength parameter controls how much impact an individual's payoff has on her reproductive success in the update step. The payoffs and selection strength can take arbitrary numerical values, but for the rest of this paper, unless otherwise noted, we will use a selection strength of $\beta = 0.5$ and the payoff matrix for this symmetric, two-player game will be:

\begin{equation}\label{eq:payoffMatrix}
\bordermatrix{&A&B&C\cr
                A& 1 &  0 & 1 \cr
                B& 0  &  2 & -4 \cr
                C& 0 & 0 & 0}
\end{equation}

Notice that in Equation \eqref{eq:payoffMatrix}, the payoff for fake news is twice the payoff for real news, but fact-checkers also inflict a stiff punishment.

\begin{figure}
    \centering
    \includegraphics[width=\textwidth]{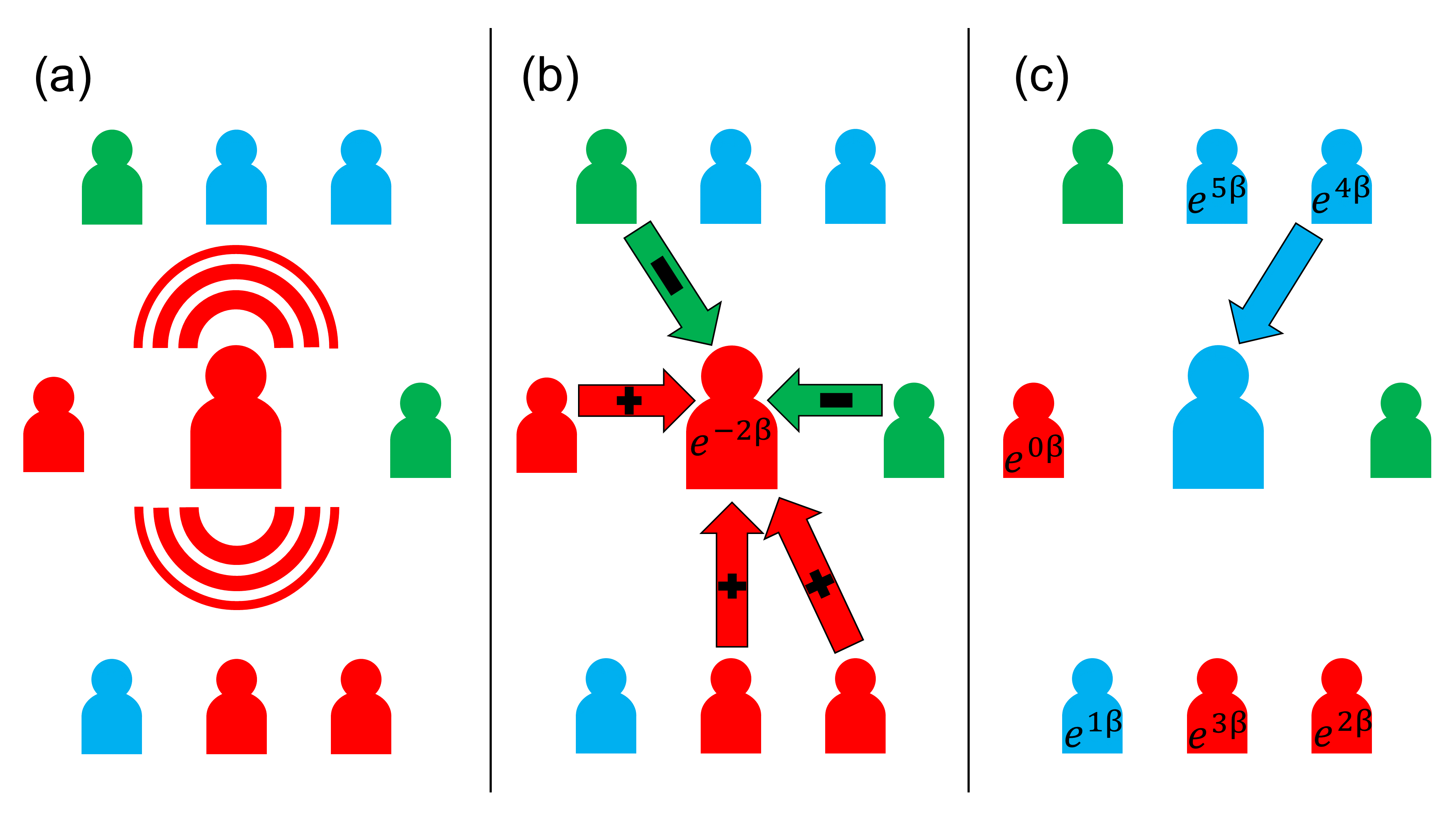}
    \caption[Fake news model schematic]{Model schematic. We model information sharing and fact-checking through the lens of spatial games. First, every individual shares news that is either true (blue) or false (red), shown in (a). In (b), we see a focal individual receiving positive or negative feedback from her neighbors depending on their relative beliefs. These information sharing dynamics are modulated by the presence of crowdsourced fact-checkers (green), characterized by the effect of their policing (positive or negative) and their static nature.   Finally, in (c), individuals in the spreading layer updates their strategy by copying a neighbor proportional to fitness.}
    \label{fig:modelCartoon}
\end{figure}

Over time, if individuals see that only certain types of stories receiving positive feedback, they may be convinced of the accuracy of those (potentially false) narratives~\cite{pennycook_cannon_rand_2018} and begin sharing those same stories themselves (Figure \ref{fig:modelCartoon}c). We will use a death-birth process for the evolutionary strategy update~\cite{ohtsuki_hauert_lieberman_nowak_2006} to capture this social imitation phenomenon. After computing the expected payoff $\pi_i$ for every individual $i$, a focal individual imitates the strategy of one of its neighbors, chosen with probability proportional to their fitness $f_i = \exp(\beta \pi_i)$. Thus, individuals with high payoff are likely to be selected, but even individuals with a low payoff due to repeated fact-checks or social isolation could be chosen to reproduce occasionally.

In our investigation, we use two flavors of this update rule: synchronous and asynchronous. In the synchronous update, used in our simulations, every individual simultaneously updates their strategy, while in the asynchronous update, which lends itself to easier mathematical analysis, a single individual is chosen uniformly at random to update. These two update rules will lead to very similar outcomes, and the minor differences between them are manifested only in edge cases that occur rarely. Keeping this in mind, we will treat them as the qualitatively same process operating on different time scales.

The basic operating procedure for our model is shown in Figure \ref{fig:modelCartoon}. First, individuals play the fake news game with neighbors by broadcasting a real or fake post. This post generates positive and negative feedback, which is converted into a fitness. Figure \ref{fig:modelCartoon}c demonstrates the asynchronous update, where only a single focal individual updates strategy by considering the fitness of all neighbors.

Our study of the spread of fake news is focused on three types of networks: a $30 \times 30$ square lattice~\cite{nowak_may_1992}, Watts-Strogatz small-world networks \cite{watts_strogatz_1998} (also with $N=900$), and a portion of the Twitter follower network \cite{twitter_data} ($N = 404719$). Our small-world networks are calibrated to have high clustering coefficients and short path lengths. To accomplish this, we use the following parameters: base degree is $8$ and the rewiring probability is $0.03,$ giving us approximately 200 shortcuts. The Twitter network is interesting for its size but also its natural clustering and the gatekeeping individuals that control the flow of information through the network. Although edges in the network were originally directed, we symmetrized the network to match the bidirectional flow of information in our model.

To initialize the system, we assign some fraction $p_C$ of the individuals as fact-checkers, and the rest we set to be A or B players with probability $\frac{1}{2}$. After initializing the system, we allow it to evolve using the one of the updating processes described above until all possible players are sharing the same type of news or a predetermined number of time steps is reached. At the end of the simulation, the type of news with more sharers is said to be dominant, and if there are no individuals sharing one type of news, we say that that strategy has gone extinct and the other strategy has fixated.

\section{Results}

\subsection{Echo Chambers and Critical Fact-checker Density}

When there are very few fact-checkers, the natural advantage that fake news sharers have allows them to drive the real news sharing strategy to extinction. Similarly, when there is a sufficient fact-checker presence, the risk of punishment for spreading misinformation is too great and the entire population shares real news. However, there is a wide range of fact-checker densities where we see the spontaneous formation of echo chambers in our simulations. We define echo chambers by their longevity, as either real or fake news goes extinct unless the minority strategy manages to form small, highly connected communities that are secured from invasion by the majority strategy. For additional discussion about the longevity of these pseudo-steady states, see the Supplementary Information. Figure \ref{fig:echoChambers} shows examples of these echo chambers on the three different network topologies we studied.

\begin{figure}
    \centering
    \includegraphics[width = \textwidth]{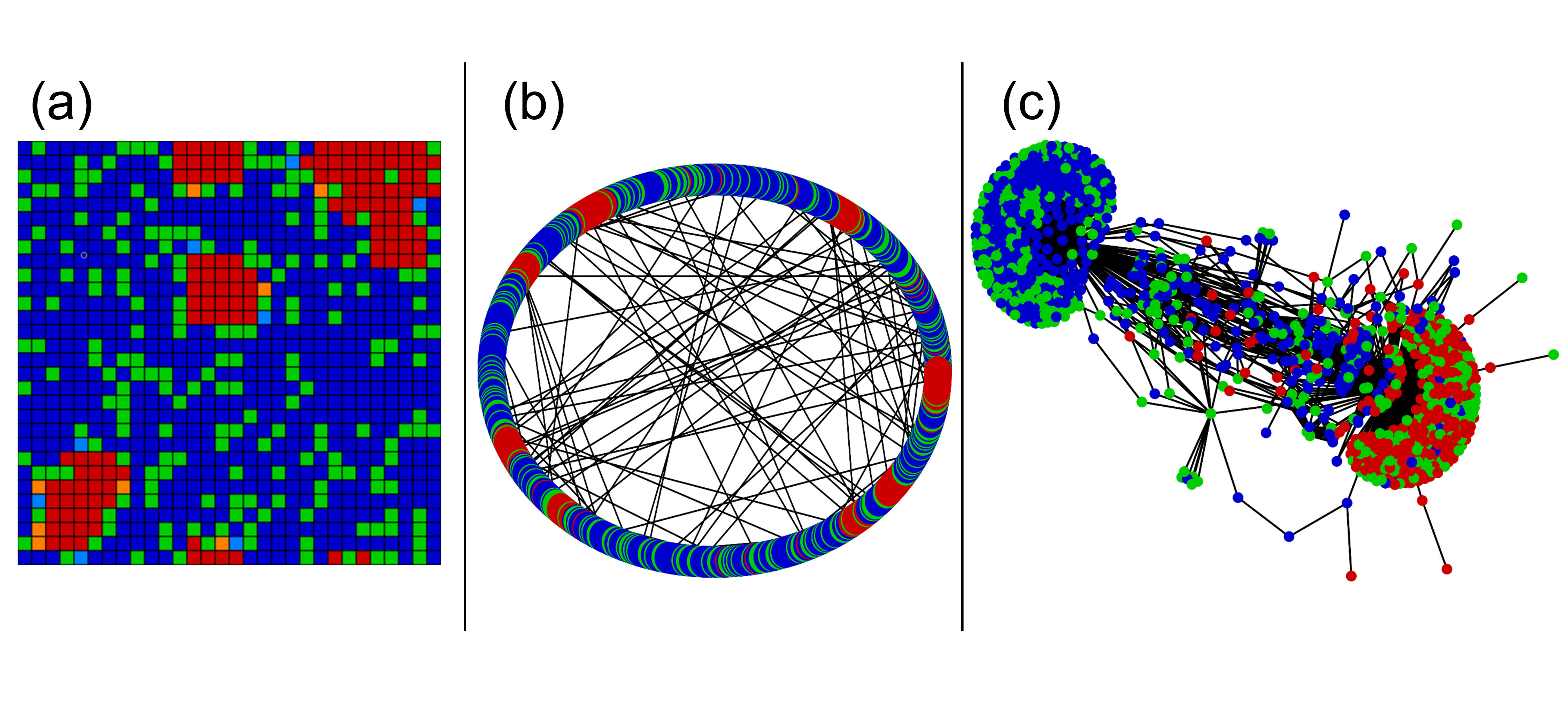}
    \caption[Echo chambers in multiple networks]{Echo chambers of fake news spreaders in a majority real news-spreading population that are isolated from the rest of the population.
    In (a), the lightly-colored individuals are those that have changed strategy recently. The network in (c) is a small breadth-first subgraph of the Twitter network of approximately 1,000 vertices, but the simulation was run on the entire $\approx 400,000$ vertex network (see Methods \& models). 
    }
    \label{fig:echoChambers}
\end{figure}

Once they form, these echo chambers are incredibly resistant to invasion, resulting in a \emph{pseudo-steady state} that cannot last forever, but will take an extremely long time to break down. After forming in a less than 100 time steps, these echo chambers remained largely unchanged for over one million time steps in our longest simulations. There may be small variation in the pseudo-steady state when specific individuals change behavior, but as a whole the echo chamber remains unchanged. Observe in Figure \ref{fig:echoChambers}a that the only individuals changing strategy are on the borders of the echo chambers in the system. It is very unlikely that a small perturbation on the border will result in any change to the interior of the echo chamber. Individuals on the periphery of the echo chamber are exposed to both views and may change strategy occasionally. Those in the interior are surrounded by like-minded individuals and have high fitness, which allows them to reinforce minority behavior by the more exposed peripheral individuals.

\begin{figure}
    \centering
    \includegraphics[width = \textwidth]{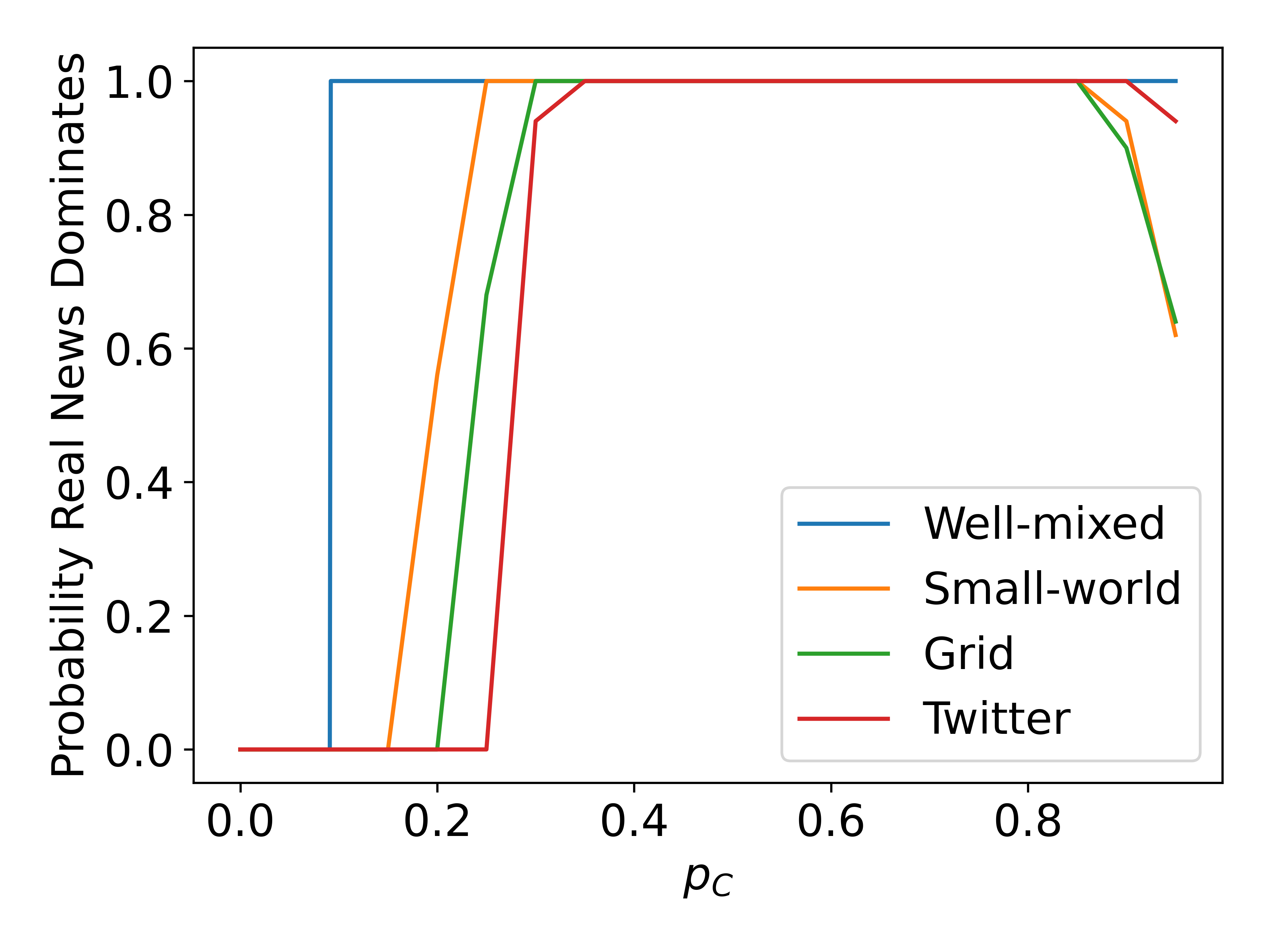}
    \caption[Critical fact-checker densities on various networks]{The probability that over half the viable population ends up sharing real news as a function of fact-checker density for different network topologies. For very high values of $p_C$, the spreader layer breaks apart into isolated individuals, at which point the dominant strategy is determined more by the random initialization than selection.}
    \label{fig:critDens}
\end{figure}

Unsurprisingly, the density of fact-checkers determines which type of news sharing is the majority and which is the minority, trapped in small and isolated communities. Figure \ref{fig:critDens} shows how the long-term behavior of the system changes as the density of fact-checkers grows. When a critical number of fact-checkers is reached, the probability of success for real news sharing increases dramatically. This critical density is different for different network types: $p_c \approx 0.235$ on the square lattice, $p_C \approx 0.2$ for small-worlds, and $p_C \approx 0.275$ for the Twitter network. These results come from simulating 50 populations at 20 different, evenly spaced fact-checker densities. At 5,000 time steps, a pseudo-steady state was declared, except the Twitter simulations which ended at 500 time steps for computational reasons.

As a note, observe that for very high values of $p_C$, the probability that real news dominates actually decreases. This seemingly paradoxical result can be explained by noting that for such high values of $p_C$, the spreader layer of the network has completely broken down into small, disconnected components. These components typically have only one or two individuals, and therefore completely constrained by their initial conditions. Selection cannot help individuals select more beneficial strategies if there are no neighbors to copy. Fortunately, when selecting fact-checkers randomly, this only occurs with unrealistically high values of $p_C$.

We can compare these results to the simple case of an infinite, well-mixed population evolving according to replicator dynamics \cite{hofbauer_sigmund_2003}. After initializing with some fraction $p_C$ of fact-checkers and the rest of the population evenly divided between real and fake news (so $p_A = p_B = \frac{1-p_C}{2}$), we consider the relative payoffs of $A$ and $B$ players when choosing a random opponent under the payoff matrix~\eqref{eq:payoffMatrix}.

The expected payoff for an $A$ player is 
\begin{equation}
f_A = 1(p_A) + 1(p_C) = \frac{1-p_C}{2} + p_C = \frac{1+p_C}{2}
\end{equation} 
and the expected payoff for a $B$ player is
\begin{equation}
    f_B = 2(p_B)-4(p_C) = 2\frac{1-p_C}{2} -4p_C = 1-5p_C
\end{equation}

Because this is a coordination game, if $A$ has a higher initial fitness, the proportion of $A$ players will grow and $f_A$ will get only get larger while $f_B$ gets smaller, until $B$ becomes functionally extinct. Therefore, the fixation of $A$ is favored over $B$ if $f_A>f_B$, which can be solving for $p_C$ using the equations above. We get the critical threshold for $p_C$, that is 
\begin{equation}
p_C> \frac{1}{11}.
\end{equation}

We conclude that the network structure of the spatial game makes containing fake news significantly more challenging. In fact, between two to three times as many fact-checkers are needed to contain the sharing of fake news in small, isolated echo chambers, and even more fact-checkers are needed to have a good chance of driving fake news sharing behavior to total extinction.

Our comprehensive simulations using the payoff values in Equation \eqref{eq:payoffMatrix} and the synchronous update rule confirm that the formation of echo chambers occurs across a wide range of payoff values, selection strengths, and network structures. Local variation in fact-checker density means in some areas there are no fact-checkers (leaving room for a fake news echo chamber) and in others they make a fact-checking wall which becomes more and more difficult for fake news sharers to penetrate as selection strength grows.

\subsection{Targeted Fact-checking}

\begin{figure}
    \centering
    \includegraphics[width = \textwidth]{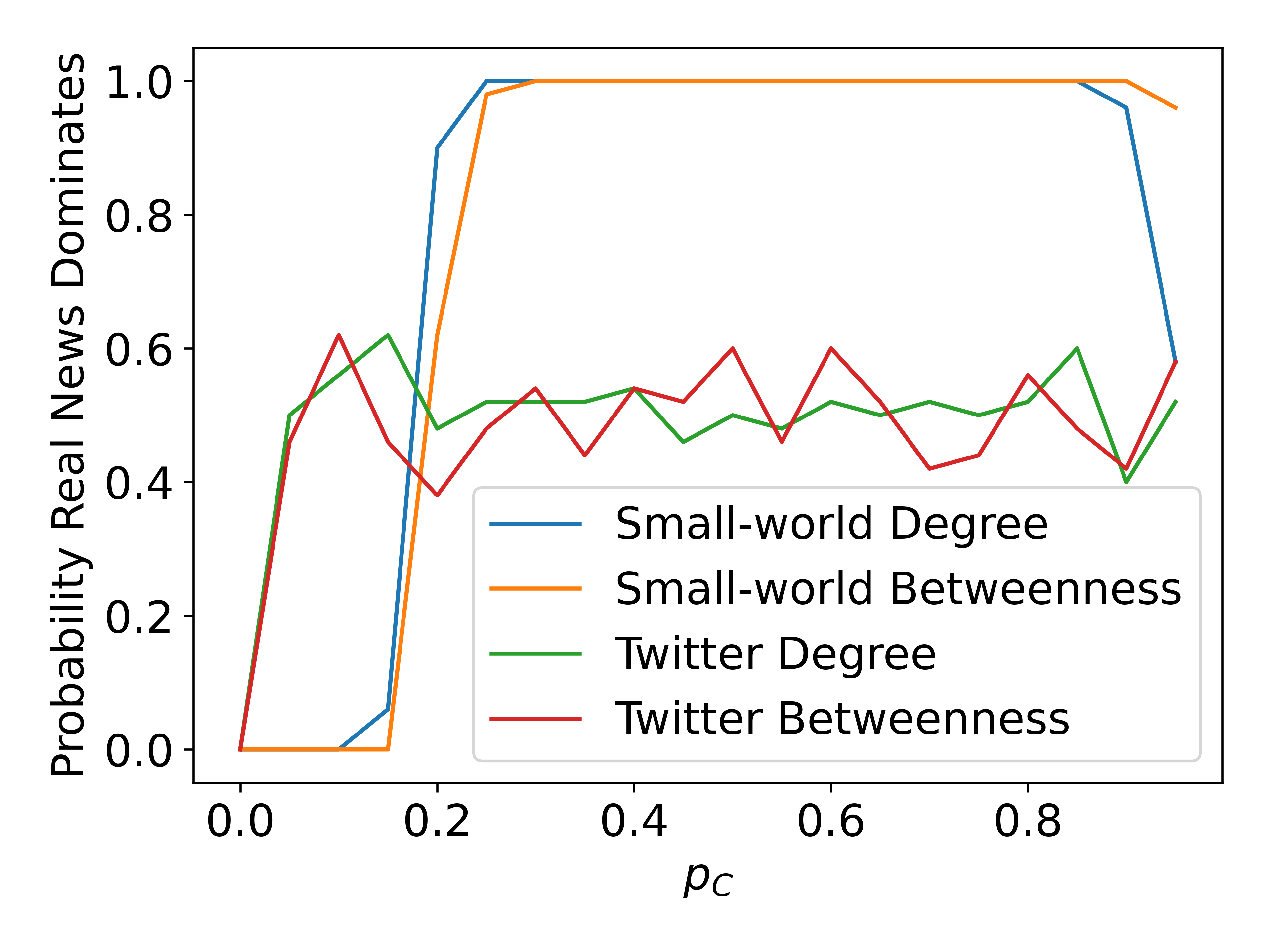}
    \caption[Effect of targeted fact-checking]{The probability of real news dominating on small-world networks and the Twitter network using the degree and betweenness centralities to place fact-checkers.}
    \label{fig:targeted}
\end{figure}

So far, we have only considered populations where fact-checkers are placed randomly. However, in almost all networks, some vertices are more centrally located than others, and this effect is particularly pronounced in naturally formed social networks. To improve the effectiveness of crowdsourced fact-checking with limited resources, it is vitally important to study targeted intervention algorithms by selecting the most central vertices. Once again, we run simulations with 50 iterations, 20 density values, and a limit of 5,000 (or 500) time steps. Our results, shown in Figure \ref{fig:targeted}, focus on two measures of network centrality, degree and betweenness \cite{barthelemy_2004}, but there are many more centrality measures and the problem of selecting individuals for optimal fact-checking remains an open problem. Since all vertices in an infinite square lattice have the same centrality, our work here is restricted to small-world networks and the Twitter network.

Figure \ref{fig:targeted} has several interesting features. First, we see that in small-worlds, using the degree and betweenness centralities have virtually the same performance. This is unsurprising as the additional shortcut edges are what create short path lengths and therefore give those individuals a high betweenness value, so the two centralities are highly correlated. More surprising is the fact that targeted fact-checking is only marginally more successful than random fact-checker placement, which can be seen by comparing Figures \ref{fig:critDens} and \ref{fig:targeted}. This may be due to the relatively uniform nature of small-world networks, where there is very little variation in degree from vertex to vertex.

However, the Twitter network has much more diversity in its degree distribution and here we see a large change between random and targeted fact-checking. By targeting high degree or betweenness centrality individuals to be fact-checkers, we quickly separate the spreader layer of the network into disconnected singletons and pairs, as these types of networks become disconnected very quickly when vertices with high degree are removed from the network \cite{albert_jeong_barabasi_2000}. Therefore, it is about equally likely that the initial random distribution will have more fake or real news sharers, so the probability that real news ``dominates'' by being present in over half the viable population hovers around 0.5 for almost all values of fact-checker density. We observed this effect for high values of $p_C$ in Figure \ref{fig:critDens}.

This suggests that in real world networks, a \emph{targeted} crowdsourced fact-checking effort where fact-checkers are also encouraged to share real news with their neighbors could be highly effective with relatively little collective effort, as the network structure will actually benefit real news instead of fake news by removing important vertices that fake news needs to move through to get to the rest of the population, while still allowing real news to spread. Enhancing our model by allowing fact-checkers to ``pass along'' real news between neighbors is one way to more effectively study targeted fact-checking algorithms.

\subsection{Analytic Results under Weak Selection} 

The selection strength $\beta$ determines the effect payoff from the fake news game has on reproductive success. As $\beta$ approaches zero \cite{nowak_sasaki_taylor_fudenberg_2004,ohtsuki_hauert_lieberman_nowak_2006}, the evolution of the system comes to resemble \emph{neutral drift}, in which individuals choose strategy with no regard for payoff. In this domain, the pseudo-steady state with its echo chambers becomes transient and short-lived. In the following section, we derive analytical results in this limit of weak selection.

Assuming a $k$-regular network structure like the square lattice, we will use an extended pair approximation method~\cite{khoo_fu_pauls_2018} to study the emergence and spread of honest behavior. The fixation probability of $A$  is the probability that a population with some initial condition evolves so that the entire viable population eventually evolves to play $A$, and we derive a closed-form expression for this probability in this work. Our aim here is to study the effects of changing the payoffs for real news, fake news, and fact-checkers, so we will begin with a general payoff matrix:

\begin{equation}\label{eq:payoffMatrixGeneral}
\bordermatrix{&A&B&C\cr
                A& a &  b & \alpha \cr
                B& c  &  d & \gamma \cr
                C& 0 & 0 & 0}
\end{equation}

In the limit of weak selection $\beta \ll 1$, we will obtain closed-form analytical conditions for the fixation probabilities of $A$ and $B$ as functions of these payoff values. 

When we suppose that we begin with a fraction $p$ of $A$ individuals, we can calculate the expected value $m_A(p)$ and variance $v_A(p)$ of the change in abundance of $A$ during the asynchronous update step where a single random individual considers changing strategy. The fixation probability of $A$ for an initial fraction $p$ of $A$ players, denoted $\rho_A(p)$, satisfies the diffusion approximation equation for large populations (see \cite{ohtsuki_hauert_lieberman_nowak_2006} for details): 
\begin{equation}
m_A(p) \frac{d}{dp} \rho_A(p) + \left( \frac{v_A(p)}{2}\right) \frac{d^2}{dp^2}\rho_A(p) = 0
\end{equation}
with the boundary conditions $\rho_A(0) =  0$ and $\rho_A(1) =  1$. This equation has closed-form solution, and thus we can obtain an exact formula for $\rho_A$. 

Our derivation of the following explicit expressions for the fixation probabilities in terms of the payoff values, lattice degree $k$, and fact-checker density $p_c$, is detailed in the Supplementary Information. For small values of $p$:

\begin{equation}\label{eq:rhoa}
    \rho_A(p) \approx p + \frac{\beta N p(1-p)}{6k}(-u_1-3u_2)
\end{equation}

\begin{equation}\label{eq:rhob}
    \rho_B(p) \approx p + \frac{\beta N p(1-p)}{6k}(-w_1-3w_2)
\end{equation}
where $u_1 = (a-b-c+d)\Big(1-k^2-\frac{1+k}{(p_C-1)(1-p_C)}\Big)$, $u_2 = -a+b+c-d-ak+bk-bk^2+dk^2+(k-1)\Big(c+(b-\alpha+\gamma)k-d(1+k)\Big)p_C$, $w_1=u_1$, and $w_2 = -(u_1+u_2)$.

In particular, we are interested in the emergence of new behavior in a previously homogeneous population. We calculate the fixation probability $\rho_A$ of a single initial $A$ player, called the invasion probability, and derive the conditions for truthful behavior to be favored, that is, when $\rho_A > 1/N$ where $N$ is the size of the population. We also repeat the process for a single $B$ player. Using Equations \eqref{eq:rhoa} and \eqref{eq:rhob}, we examine the effect $p_C$ and $\gamma$, the punishment defectors suffer from fact-checkers, have on the invasion probabilities of real and fake news.

\begin{figure}
    \centering
    \includegraphics[width=\textwidth]{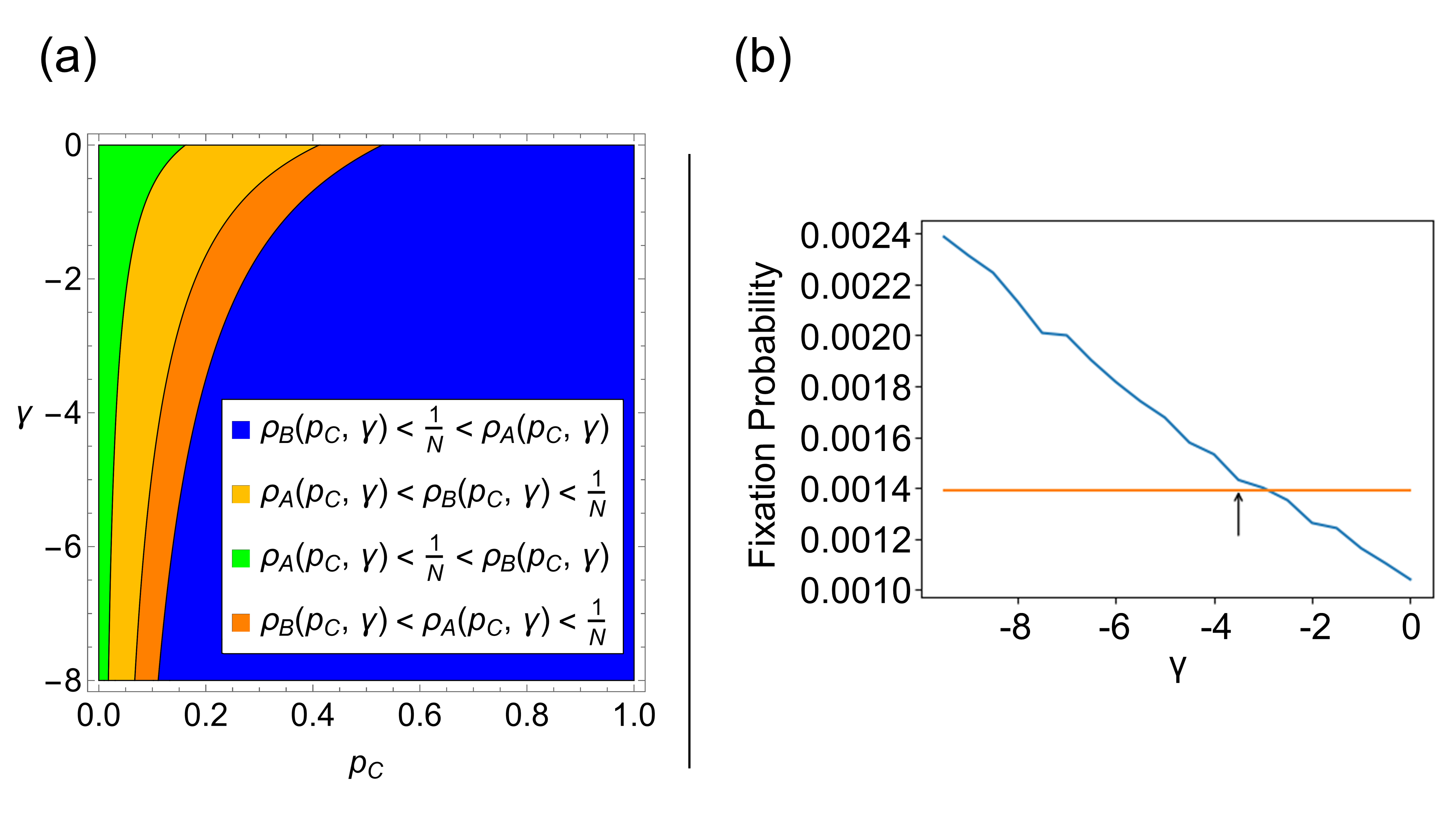}
    \caption[Invasion probabilities under weak selection]{The invasion probabilities of real and fake news spreaders in the limit of weak selection using payoff values from \eqref{eq:payoffMatrix}, except for $\gamma$ which varies from $0$ to $-8$. In (a), we see what regions of the $p_C - \gamma$ plane give true stories an advantage (blue region), false news an advantage (green region), or neither (orange regions). In (b), we see an approximation of the invasion probability for a single real news sharer from simulations, when $p_C = 0.2$ and $\beta = 0.0001$. These simulation results intersect the threshold line $\frac{1}{N} \approx 0.0014$ close to where it was predicted by the analytic results, indicated by the arrow.}
    \label{fig:weakSelectionFixation}
\end{figure}

This allows us to determine the conditions under which fact-checking will be effective at stemming misinformation and quantify how steep the penalty $\gamma$ needs to be for a given proportion of fact-checkers, $p_C$, in the system. In Figure \ref{fig:weakSelectionFixation}a, we see that for strong penalties, $\gamma < -4$, only a fifth of the population or less needs to be fact-checkers for selection to favor real news. However, as $\gamma$ gets closer to zero, the number of fact-checkers need goes up to about half the population. The green region of the $p_C-\gamma$ plane shows where selection favors fake news; this only happens when there are very few fact-checkers. Notice that there is a wide region in orange where selection does not favor invasion by real or fake news. This is because the fake news game is a coordination game that tends to put minorities (like a single invading mutant) at a disadvantage. These analytic approximations closely match extensive simulations, as shown in Figure \ref{fig:weakSelectionFixation}b.

\section{Discussion and Conclusion}

This work adds to the growing body of research surrounding fake news, echo chambers, and fact-checking and we believe that this work has immediate implications for the study of misinformation. We have shown that the spatial structure of social networks tends to favor the spread of fake news, but by carefully selecting fact-checkers, that same structure can be used to combat misinformation by amplifying the effects of fact-checking. 

Our analytic results allow us to easily test potential combinations of reward and punishment and use both ``carrots and sticks'' to encourage real news and dampen fake news. Like previous work studying public goods games, we see that a strong punishment of defectors is effective at stopping bad behavior~\cite{sigmund_hauert_nowak_2001, sigmund_desilva_traulsen_hauert_2010, helbing_szolnoki_perc_szabo_2010}.

Future work combining potential experimental behavior data~\cite{pennycook_cannon_rand_2018} with our present model will help incorporate relevant social network and psychological factors in our research. In particular, the constants in the payoff matrix and the selection strength were chosen fairly arbitrarily. Analyzing real-world data may allow us better estimates of some of these values, which in turn can give better actionable advice about how to actually control the spread of fake news. We would also like to analyze preexisting data sets or create new empirical studies to confirm our predictions regarding the effects that the rewards and punishments of sharing real and fake news have on the ability of fake news to spread through a population. As an example, perhaps placing fact-checking comments at the top of any fake news threads would sufficiently increase the punishment suffered by fake news's sharers to prevent its spread. 

Recent theoretical research has demonstrated that partisan bias~\cite{kawakatsu_lelkes_levin_tarnita_2021} and information cascades~\cite{tokita_guess_tarnita_2021} are two possible explanations for the formation of echo chambers. Our work here shows that the spatial distribution of fact-checkers can contribute to echo chamber creation. This work only represents the first steps towards understanding how fact-checkers impact echo chamber formation. These echo chambers require certain conditions to form, including an appropriate selection strength, but there is much we still do not understand. Preliminary results show that the formation of resilient echo chambers is dependent on the type of network used. While social media sites do resemble lattices or small worlds in some respects, there are other properties of social networks that may be more or less conducive to echo chamber formation.

Extensions of our present work on targeted fact-checking efforts will likely lead to useful insights for optimizing field deployment of crowdsourcing fact-checking.  There will be a good deal of further work to do, for example, on using other network topologies and other targeting centralities. In addition, the use of larger network data sets will give us more realistic behavior as there may be large-scale social network features essential to the development of echo chambers that are not captured in any of the network models we used.

Last but not least, our present work will help stimulate future work extending targeting algorithms to multiplex networks that take into account the fact that the interconnected ecosystems of social media platforms enable multi-channel communication and spillover from one platform to the other. In doing so, we hope to develop mechanistic models that allow us to explore realistic extensions incorporating social psychological factors such as heterogeneity of social influence, repeated exposure, and pre-existing beliefs.

\section{References}
\medskip

\bibliography{newrefs}

\begin{thebibliography}{10}

\bibitem{lazer_baum_benkler_berinsky_greenhill_menczer_metzger_nyhan_pennycook_rothschild_schudson_sloman_sunstein_thorson_watts_zittrain_2018}
David M.~J. Lazer, Matthew~A. Baum, Yochai Benkler, Adam~J. Berinsky, Kelly~M.
  Greenhill, Filippo Menczer, Miriam~J. Metzger, Brendan Nyhan, Gordon
  Pennycook, David Rothschild, Michael Schudson, Steven~A. Sloman, Cass~R.
  Sunstein, Emily~A. Thorson, Duncan~J. Watts, and Jonathan~L. Zittrain.
\newblock The science of fake news.
\newblock {\em Science}, 359(6380):1094--1096, 2018.

\bibitem{shin_jian_driscoll_bar_2018}
Jieun Shin, Lian Jian, Kevin Driscoll, and François Bar.
\newblock The diffusion of misinformation on social media: Temporal pattern,
  message, and source.
\newblock {\em Computers in Human Behavior}, 83:278–287, 2018.

\bibitem{vosoughi_roy_aral_2018}
Soroush Vosoughi, Deb Roy, and Sinan Aral.
\newblock The spread of true and false news online.
\newblock {\em Science}, 359(6380):1146–1151, 2018.

\bibitem{wang_sirianni_tang_zheng_fu_2020}
Xin Wang, Antonio~D. Sirianni, Shaoting Tang, Zhiming Zheng, and Feng Fu.
\newblock Public discourse and social network echo chambers driven by
  socio-cognitive biases.
\newblock {\em Physical Review X}, 10(4):041042, 2020.

\bibitem{poole_rosenthal_1985}
Keith~T. Poole and Howard Rosenthal.
\newblock A spatial model for legislative roll call analysis.
\newblock {\em American Journal of Political Science}, 29(2):357--384, 1985.

\bibitem{delvicario_bessi_zollo_petroni_scala_caldarelli_stanley_quattrociocchi_2016}
Michela Del~Vicario, Alessandro Bessi, Fabiana Zollo, Fabio Petroni, Antonio
  Scala, Guido Caldarelli, H.~Eugene Stanley, and Walter Quattrociocchi.
\newblock The spreading of misinformation online.
\newblock {\em Proceedings of the National Academy of Sciences},
  113(3):554–559, 2016.

\bibitem{mian_khan_2020}
Areeb Mian and Shujhat Khan.
\newblock Coronavirus: The spread of misinformation.
\newblock {\em BMC Medicine}, 18(1):89, 2020.

\bibitem{bursztyn_rao_roth_yanagizawa-drott_2020}
Leonardo Bursztyn, Aakaash Rao, Christopher Roth, and David Yanagizawa-Drott.
\newblock Misinformation during a pandemic.
\newblock Working Paper 27417, National Bureau of Economic Research, 2020.

\bibitem{pennycook_mcphetres_zhang_lu_rand_2020}
Gordon Pennycook, Jonathon McPhetres, Yunhao Zhang, Jackson~G. Lu, and David~G.
  Rand.
\newblock Fighting {COVID}-19 misinformation on social media: Experimental
  evidence for a scalable accuracy-nudge intervention.
\newblock {\em Psychological Science}, 31(7):770–780, 2020.

\bibitem{andris_lee_hamilton_martino_gunning_selden_2015}
Clio Andris, David Lee, Marcus~J. Hamilton, Mauro Martino, Christian~E.
  Gunning, and John~Armistead Selden.
\newblock The rise of partisanship and super-cooperators in the {U.S.} house of
  representatives.
\newblock {\em PLOS ONE}, 10(4):1--14, 2015.

\bibitem{pewreseachcenter_2017b}
Pew~Research Center.
\newblock Partisan conflict and congressional outreach.
\newblock Technical report, Pew Research Center, February 2017.

\bibitem{pewreseachcenter_2014}
Pew~Research Center.
\newblock Political polarization in the {American} public.
\newblock Technical report, Pew Research Center, June 2014.

\bibitem{pewreseachcenter_2016a}
Pew~Research Center.
\newblock Partisanship and political animosity in 2016.
\newblock Technical report, Pew Research Center, June 2016.

\bibitem{pewreseachcenter_2016b}
Pew~Research Center.
\newblock Few {Clinton} or {Trump} supporters have close friends in the other
  camp.
\newblock Technical report, Pew Research Center, August 2016.

\bibitem{fu_wang_2008}
Feng Fu and Long Wang.
\newblock Coevolutionary dynamics of opinions and networks: From diversity to
  uniformity.
\newblock {\em Physical Review E}, 78(1):016104, 2008.

\bibitem{holme_newman_2006}
Petter Holme and M.~E. Newman.
\newblock Nonequilibrium phase transition in the coevolution of networks and
  opinions.
\newblock {\em Physical Review E}, 74(5):056108, 2006.

\bibitem{zanette_gil_2006}
Damián~H. Zanette and Santiago Gil.
\newblock Opinion spreading and agent segregation on evolving networks.
\newblock {\em Physica D: Nonlinear Phenomena}, 224(1-2):156–165, 2006.

\bibitem{nardini_kozma_barrat_2008}
Cecilia Nardini, Balázs Kozma, and Alain Barrat.
\newblock Who’s talking first? {Consensus} or lack thereof in coevolving
  opinion formation models.
\newblock {\em Physical Review Letters}, 100(15):158701, 2008.

\bibitem{friedkin_proskurnikov_tempo_parsegov_2016}
Noah~E. Friedkin, Anton~V. Proskurnikov, Roberto Tempo, and Sergey~E. Parsegov.
\newblock Network science on belief system dynamics under logic constraints.
\newblock {\em Science}, 354(6310):321–326, 2016.

\bibitem{wang_rong_wu_2016}
Shaoli Wang, Libin Rong, and Jianhong Wu.
\newblock Bistability and multistability in opinion dynamics models.
\newblock {\em Applied Mathematics and Computation}, 289:388–395, 2016.

\bibitem{antonopoulos_shang_2018}
Chris~G. Antonopoulos and Yilun Shang.
\newblock Opinion formation in multiplex networks with general initial
  distributions.
\newblock {\em Scientific Reports}, 8(1):2852, 2018.

\bibitem{stewart_arechar_rand_plotkin_2021}
Alexander~J. Stewart, Antonio~A. Arechar, David~G. Rand, and Joshua~B. Plotkin.
\newblock The coercive logic of fake news.
\newblock {\em arXiv preprint abs/2108.13687}, 2021.

\bibitem{evans_fu_2018}
Tucker Evans and Feng Fu.
\newblock Opinion formation on dynamic networks: Identifying conditions for the
  emergence of partisan echo chambers.
\newblock {\em Royal Society Open Science}, 5(10):181122, 2018.

\bibitem{perc_szolnoki_2010}
Matja\v{z} Perc and Attila Szolnoki.
\newblock Coevolutionary games—a mini review.
\newblock {\em Biosystems}, 99(2):109–125, 2010.

\bibitem{schmidt_zollo_scala_betsch_quattrociocchi_2018}
Ana~Lucía Schmidt, Fabiana Zollo, Antonio Scala, Cornelia Betsch, and Walter
  Quattrociocchi.
\newblock Polarization of the vaccination debate on {Facebook}.
\newblock {\em Vaccine}, 36(25):3606–3612, 2018.

\bibitem{ohtsuki_hauert_lieberman_nowak_2006}
Hisashi Ohtsuki, Christoph Hauert, Erez Lieberman, and Martin~A. Nowak.
\newblock A simple rule for the evolution of cooperation on graphs and social
  networks.
\newblock {\em Nature}, 441(7092):502–505, 2006.

\bibitem{tarnita_ohtsuki_antal_fu_nowak_2009}
Corina~E. Tarnita, Hisashi Ohtsuki, Tibor Antal, Feng Fu, and Martin~A. Nowak.
\newblock Strategy selection in structured populations.
\newblock {\em Journal of Theoretical Biology}, 259(3):570–581, 2009.

\bibitem{nowak_may_1992}
Martin~A. Nowak and Robert~M. May.
\newblock Evolutionary games and spatial chaos.
\newblock {\em Nature}, 359(6398):826–829, 1992.

\bibitem{pennycook_rand_2019}
Gordon Pennycook and David~G. Rand.
\newblock Fighting misinformation on social media using crowdsourced judgments
  of news source quality.
\newblock {\em Proceedings of the National Academy of Sciences},
  116(7):2521–2526, 2019.

\bibitem{roozenbeek_vanderlinden_2019a}
Jon Roozenbeek and Sander van~der Linden.
\newblock The fake news game: Actively inoculating against the risk of
  misinformation.
\newblock {\em Journal of Risk Research}, 22(5):570–580, 2019.

\bibitem{roozenbeek_vanderlinden_2019b}
Jon Roozenbeek and Sander van~der Linden.
\newblock Fake news game confers psychological resistance against online
  misinformation.
\newblock {\em Palgrave Communications}, 5(1):65, 2019.

\bibitem{mcguire_papageorgis_1962}
William~J. McGuire and Demetrios Papageorgis.
\newblock Effectiveness of forewarning in developing resistance to persuasion.
\newblock {\em Public Opinion Quarterly}, 26(1):24--34, 1962.

\bibitem{banas_rains_2010}
John~A. Banas and Stephen~A. Rains.
\newblock A meta-analysis of research on inoculation theory.
\newblock {\em Communication Monographs}, 77(3):281–311, 2010.

\bibitem{pennycook_cannon_rand_2018}
Gordon Pennycook, Tyrone~D. Cannon, and David~G. Rand.
\newblock Prior exposure increases perceived accuracy of fake news.
\newblock {\em Journal of Experimental Psychology: General},
  147(12):1865–1880, 2018.

\bibitem{watts_strogatz_1998}
Duncan~J. Watts and Steven~H. Strogatz.
\newblock Collective dynamics of ‘small-world’ networks.
\newblock {\em Nature}, 393(6684):440–442, 1998.

\bibitem{twitter_data}
Ryan~A. Rossi and Nesreen~K. Ahmed.
\newblock The network data repository with interactive graph analytics and
  visualization.
\newblock In {\em Proceedings of the Twenty-Ninth AAAI Conference on Artificial
  Intelligence}, AAAI'15, page 4292–4293. AAAI Press, 2015.

\bibitem{hofbauer_sigmund_2003}
Josef Hofbauer and Karl Sigmund.
\newblock Evolutionary game dynamics.
\newblock {\em Bulletin of the American Mathematical Society}, 40(4):479–519,
  2003.

\bibitem{barthelemy_2004}
Marc Barthelemy.
\newblock Betweenness centrality in large complex networks.
\newblock {\em The European Physical Journal B - Condensed Matter},
  38(2):163–168, 2004.

\bibitem{albert_jeong_barabasi_2000}
R\'{e}ka Albert, Hawoong Jeong, and Albert-L\'{a}szl\'{o} Barab\'{a}si.
\newblock Error and attack tolerance of complex networks.
\newblock {\em Nature}, 406(6794):378–382, 2000.

\bibitem{nowak_sasaki_taylor_fudenberg_2004}
Martin~A. Nowak, Akira Sasaki, Christine Taylor, and Drew Fudenberg.
\newblock Emergence of cooperation and evolutionary stability in finite
  populations.
\newblock {\em Nature}, 428(6983):646–650, 2004.

\bibitem{khoo_fu_pauls_2018}
Tommy Khoo, Feng Fu, and Scott Pauls.
\newblock Spillover modes in multiplex games: Double-edged effects on
  cooperation and their coevolution.
\newblock {\em Scientific Reports}, 8(1):6922, 2018.

\bibitem{sigmund_hauert_nowak_2001}
K.~Sigmund, C.~Hauert, and M.~A. Nowak.
\newblock Reward and punishment.
\newblock {\em Proceedings of the National Academy of Sciences},
  98(19):10757–10762, 2001.

\bibitem{sigmund_desilva_traulsen_hauert_2010}
Karl Sigmund, Hannelore De~Silva, Arne Traulsen, and Christoph Hauert.
\newblock Social learning promotes institutions for governing the commons.
\newblock {\em Nature}, 466(7308):861–863, 2010.

\bibitem{helbing_szolnoki_perc_szabo_2010}
Dirk Helbing, Attila Szolnoki, Matja\v{z} Perc, and György Szabó.
\newblock Punish, but not too hard: How costly punishment spreads in the
  spatial public goods game.
\newblock {\em New Journal of Physics}, 12(8):083005, 2010.

\bibitem{kawakatsu_lelkes_levin_tarnita_2021}
Mari Kawakatsu, Yphtach Lelkes, Simon~A. Levin, and Corina~E. Tarnita.
\newblock Interindividual cooperation mediated by partisanship complicates
  {Madison’s} cure for “mischiefs of faction”.
\newblock {\em Proceedings of the National Academy of Sciences},
  118(50):e2102148118, 2021.

\bibitem{tokita_guess_tarnita_2021}
Christopher~K. Tokita, Andrew~M. Guess, and Corina~E. Tarnita.
\newblock Polarized information ecosystems can reorganize social networks via
  information cascades.
\newblock {\em Proceedings of the National Academy of Sciences},
  118(50):e2102147118, 2021.

\end{thebibliography}

\end{document}


\maketitle

This supplementary information contains some additional exploration of the fake news spatial game described in the main paper, as well as the derivation of the invasion probabilities in the limit of weak selection.

\section{Echo Chamber Longevity and the Pseudo-steady State}

\begin{figure}
    \centering
    \includegraphics[width = \textwidth]{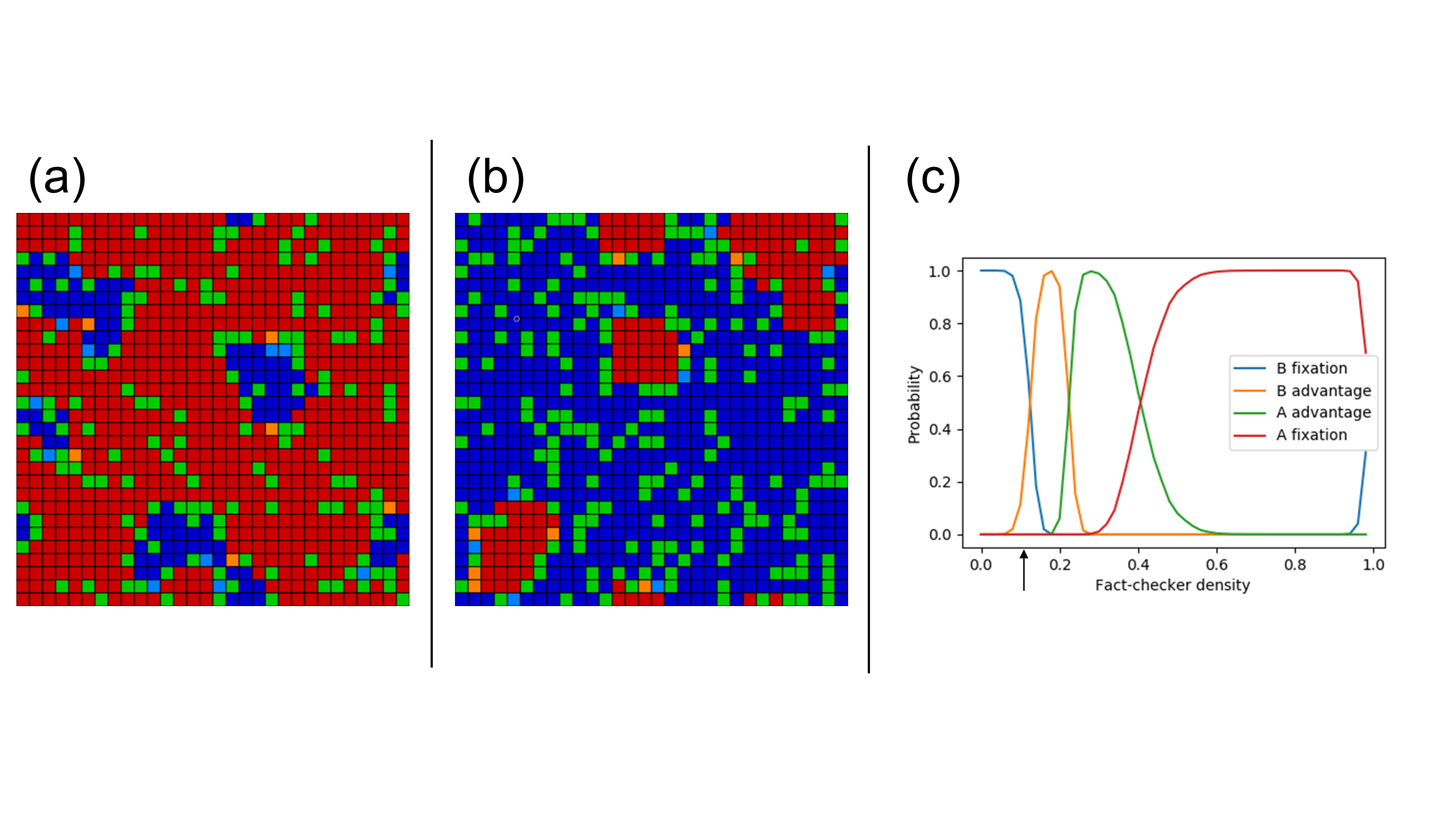}
    \caption{Panels a and b show echo chambers of real news (blue) or fake news (red) sharers that are isolated from the rest of the population by a barrier of fact-checkers (green). Lightly-colored individuals are those that have changed strategy in the last time step. The plot in (c) used simulations to show how the long-term behavior changes as the fact-checker density varies, with the arrow indicating the fact-checker density at which real news has an advantage in a well-mixed population, $p_C = 1/11$. As the number of fact-checkers increases, the population moves towards more real news and less false news stories being shared.}
    \label{fig:echoChambers}
\end{figure}

In this section, we expand our investigation into the role fact-checkers play in containing the spread of fake news. The density of static fact-checkers has a significant effect on the formation of echo chambers and which strategy ``dominates'' by controlling over half the viable population. Fig \ref{fig:echoChambers}a and \ref{fig:echoChambers}b show two examples of this on the square lattice. Different strategies dominate, dependent on the fact-checker density. 

In the main paper, we focused on a critical value of $p_C$ at which point selection favors real news instead of fake. However, there is an additional point to consider. Instead of simply containing fake news to isolated echo chambers, we may want to select enough fact-checkers to completely eradicate fake news. On the other hand, for a sufficiently small number of fact-checkers, it is extremely likely that eventually the entire population will be sharing fake news.
Therefore, there are actually four different regions of behavior: fake news (B) fixates and real news (A) goes extinct, fake news has the advantage in the population with small real news echo chambers, real news has the advantage with small fake news echo chambers, and real news fixates while fake news goes extinct. This sequence of behaviors and their probabilities are shown in Fig \ref{fig:echoChambers}c. 

We can see the formation of echo chambers for a wide range of fact-checker densities, approximately 0.15 to 0.5 in the case of the square lattice with selection strength $\beta = 0.5$. We call behavior in this region the pseudo-steady state because these echo chambers are highly resistant to invasion and thus can persist for millions of time steps. However, it is not a true steady state because with an infinite amount of time, eventually the echo chambers will break down and one strategy will go extinct.

\begin{figure}
    \centering
    \includegraphics[width = \textwidth]{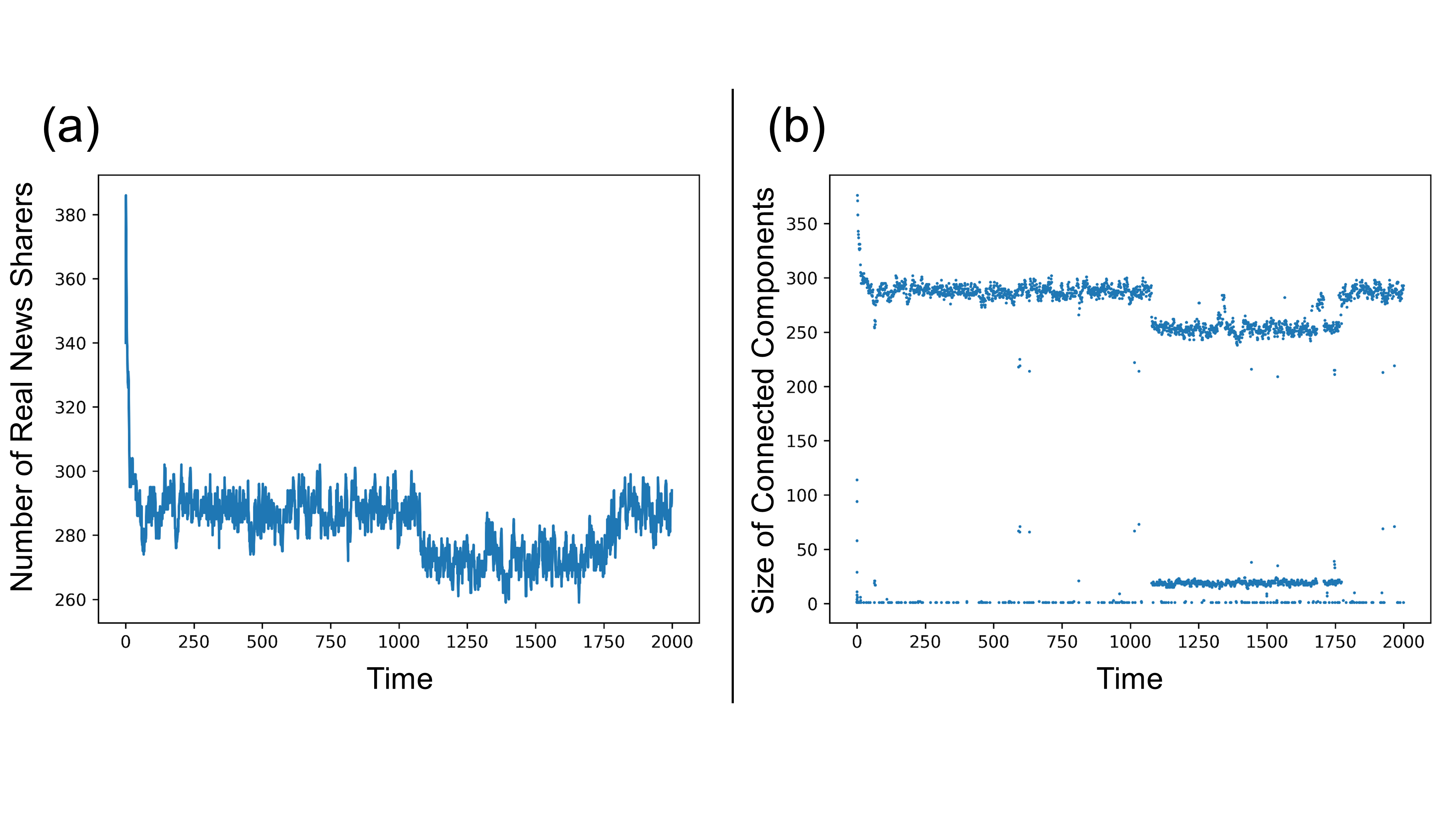}
    \caption{The characteristic evolution of a 900 individual population with $p_C = 0.2$ over the course of 2000 time steps. In (a), we can see that after a short chaotic period, the system reaches a pseudo-steady state and the number of true news sharers is fairly constant except for short bursts of disruption when clusters of individuals all shift strategy together. In (b), we get a more detailed look at what happened in the same system by looking at the size of individual connected components. Around $t=1100$, the single large component of real news sharers splits into two separate components. Then at about $t=1800$, the two components are joined together as a small cluster between them changes back to sharing real news.}
    \label{fig:clusters}
\end{figure}

We can see the resilience of these echo chambers by looking at the number of real news sharers as a function of time. Figure \ref{fig:clusters}a shows the prevalence of real news in a single representative simulation. The number of cooperators drops swiftly at first before stabilizing at around 290 cooperators. There are small shifts at $t \approx 1100$ and $t \approx 1800$, but otherwise the population is unchanging except for minor perturbations on the border of echo chambers. Fig \ref{fig:clusters}b gives more detail, showing the size of each path-connected component of real news sharers. By comparing Fig \ref{fig:clusters}a and b, we see that the changes in cooperator population size corresponds to the large 290-individual echo chamber breaking into two smaller components, one with $\approx 250$ individuals and the other with $\approx 20$, and then fusing back together.

On the square lattice, the formation of echo chambers and the pseudo-steady state seems to occur across a wide range of fact-checker densities. As shown in the main paper, we also observe echo chamber formation on small-world networks and the twitter network. However, this is not a uniform property of all networks. Preliminary results show that the formation of echo chambers and the critical $p_C$ value are dependent on network topology; lattices and small-worlds are fertile ground for echo chambers, but Erd\"{o}s-Reny\'{i} random graphs and scale-free networks are not. This leads us to hypothesize that a relatively high clustering coefficient is essential for the formation of echo chambers. This intuitively makes sense, as echo chambers are dependent on the feedback loops possible in cliquish, highly connected communities.

\section{Fact-checker Inaccuracy}

\begin{figure}
    \centering
    \includegraphics[width=\textwidth]{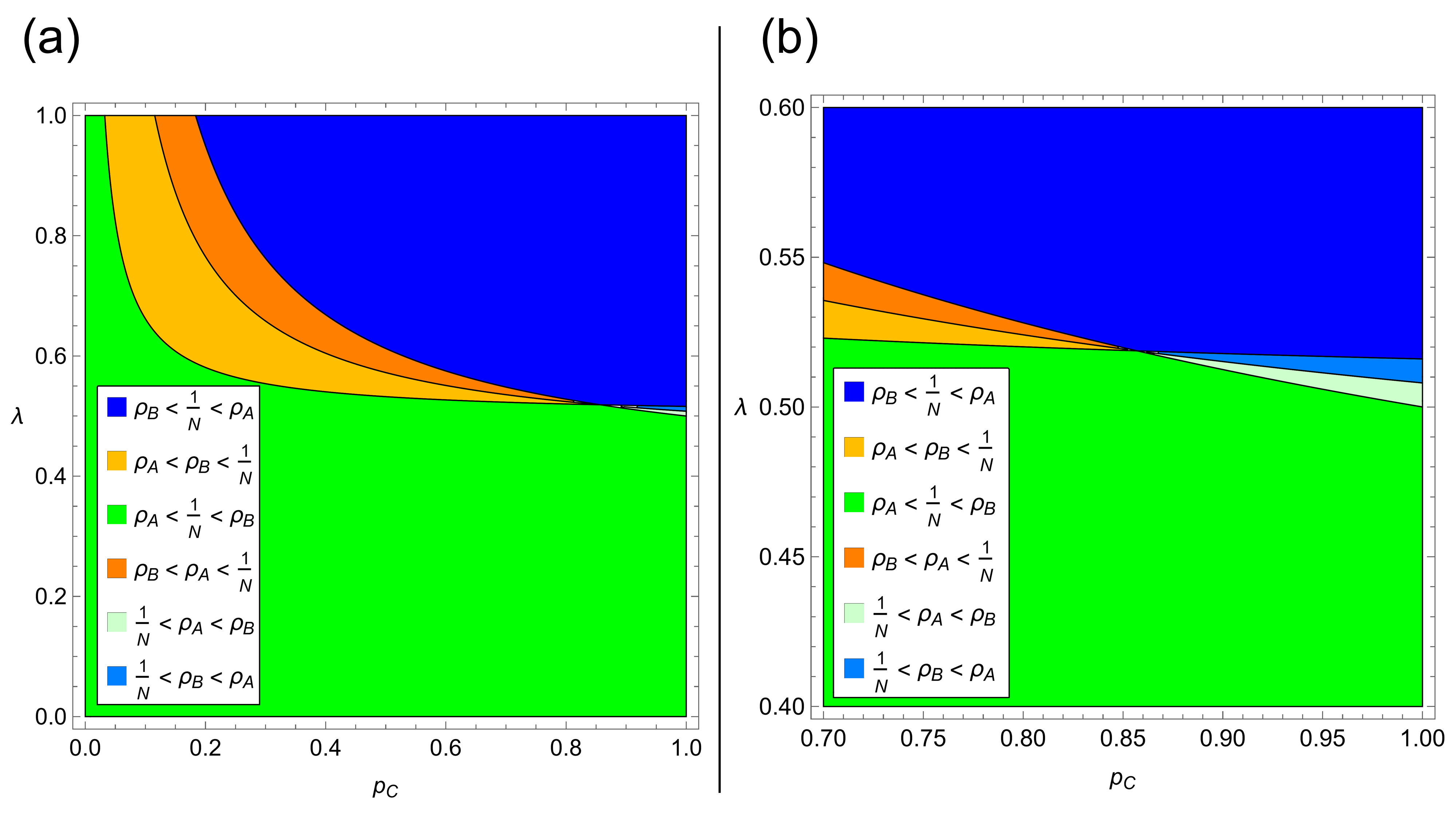}
    \caption{The results of varying the accuracy of fact-checkers. In (a), we see the where in the $p_C - \lambda$ plane selection favors true news (blue), false news (green), or neither (orange). However, when the density of fact-checkers is very high and fact-checkers are not very accurate, selection can actually favor invasion by true or false news, as shown in (b). This is surprising because this is a coordination game and it is rare for selection to favor invasion by both strategies. However, this combination of parameter values is highly unrealistic and would never occur in real life.}
    \label{fig:weakSelectionLambda}
\end{figure}

In reality, fact-checking is subject to human errors. Some fake news occasionally goes unnoticed and endorsed, and some real news is temporally labelled to be fake by well-meaning fact-checkers. When relying on citizen fact-checkers instead of professional journalists for peer policing purposes, the accuracy of fact-checking will inevitably go down as laymen are less prepared to accurately assess fake news. Suppose that fact-checkers have an accuracy in their policing of $\lambda \in [0, 1]$. With probability $\lambda$, they correctly assess a post's accuracy and reward benefit $\alpha$ to true news spreaders and penalty $\gamma$ to fake news spreaders. With probability $1-\lambda$, an error occurs, leading to the opposite payoff assignments. Using the same method we use to calculate the analytic fixation probabilities, we will quantify the precision threshold required for fact-checkers to ensure fair and transparent policing of wrongdoers while in favor of real news spreaders. For the exact expressions, see the end of the section on analytic derivations below. Figure \ref{fig:weakSelectionLambda} shows the relationship between invasion probabilities on the $p_C - \lambda$ plane when using the following payoff matrix:

\begin{equation}\label{eq:payoffMatrixSI}
\bordermatrix{&A&B&C\cr
                A& 1 &  0 & 1 \cr
                B& 0  &  2 & -4 \cr
                C& 0 & 0 & 0}
\end{equation}

In Fig \ref{fig:weakSelectionLambda}a, we see that when $\lambda < 0.5$, selection always favors fake news. This is unsurprising, as it means that the supposed fact-checkers are actually giving more benefit to fake news spreaders than real news spreaders. However, there is a clear buffer in which fact-checkers can be accurate only about 80\% of the time without necessitating a drastic increase in the critical fact-checker density for selection to favor real news. 

Fig \ref{fig:weakSelectionLambda}b shows an interesting phenomenon. When fact-checker accuracy is very close to $1/2$ and the number of fact-checkers is extremely high, selection actually favors invasion by both real \emph{and} fake news. This is surprising because this real vs fake news game is a coordination game which tends to oppose invading mutants. While this set of parameters is unrealistic and would never appear in any real population, it still demonstrates an interesting property of the dynamics of coordination games in the presence of zealots or extreme environmental conditions.

\section{Derivation of Analytic Results}

In this section, we derive the invasion probabilities of single cooperators and defectors in the limit of weak selection. We begin by introducing the necessary notation. We have $N$ individuals on a network, each with $k$ neighbors, and they play a game with a general payoff matrix
\begin{equation}
\bordermatrix{&A&B&C\cr
                A& a & b & \alpha\cr
                B& c & d & \gamma\cr
                C& 0 & 0 & 0}
\end{equation}
$p_A$, $p_B$, and $p_C$ are the proportions of $A$, $B$, and $C$ players. Similarly, $p_{S_1 S_2}$ is the proportion of edges leading from an individual playing $S_1$ to an individual playing $S_2$, where $S_1$ and $S_2$ can be $A$, $B$, or $C$. We will also be interested in the conditional probability of finding an individual playing $S_2$ by following a random edge that starts at an individual playing $S_1$, which will be denoted $q_{S_2|S_1}$. By basic probability, $q_{S_2|S_1} = \frac{p_{S_1 S_2}}{p_{S_1}}$. 

For an individual playing $S_i$, $\pi_{S_i}$ is the total payoff, or the sum of the payoffs from each interaction with a neighbor. The payoff of any $A$ or $B$ individual is dependent on the neighbors' strategies, but we are interested in the expected payoff which only depends on the quantities already listed. With selection strength $\beta$, $f_{S_i} = e^{\beta\pi_{S_i}}$ is the fitness of an individual playing $S_i$.

We have two normalization conditions that ensure that all our probabilities sum to 1:
\begin{equation} p_A + p_B + p_C = 1 \end{equation}
\begin{equation} p_{AA} + p_{AB} + p_{AC} + p_{BA} + p_{BB} + p_{BC} + p_{CA} + p_{CB} + p_{CC} = 1 \end{equation}

Additionally, there are three symmetry conditions. These need not be true in general, but because the network we are using is undirected, an edge from $S_1$ to $S_2$ is also an edge from $S_2$ to $S_1$. Therefore:
\begin{equation} p_{AB} = p_{BA}\end{equation}
\begin{equation} p_{AC} = p_{CA} \end{equation}
\begin{equation} p_{BC} = p_{CB} \end{equation}

Finally, we have three consistency conditions:
\begin{equation} p_A = p_{AA} + p_{AB} + p_{AC} \end{equation}
\begin{equation} p_B = p_{BA} + p_{BB} + p_{BC} \end{equation}
\begin{equation} p_C = p_{CA} + p_{CB} + p_{CC} \end{equation}

With all these conditions, we can simplify the system until there are only five independent variables: $p_A$, $p_B$, $p_{AA}$, $p_{BB}$, $p_{CC}$. The other four variables can be solved in terms of these five:
\begin{equation} p_C = 1 - p_A - p_B \end{equation}
\begin{equation} \label{eq:pab}
 p_{AB} = p_{BA} = 1/2\Big[(p_A-p_{AA})+(p_B-p_{BB})-(p_C-p_{CC})\Big]\end{equation}
\begin{equation}\label{eq:pac}
p_{AC} = p_{CA} = 1/2\Big[(p_C-p_{CC})-(p_B-p_{BB})+(p_A-p_{AA})\Big] \end{equation}
\begin{equation} \label{eq:pbc}
p_{BC} = p_{CB} = 1/2\Big[(p_B-p_{BB})+(p_C-p_{CC})-(p_A-p_{AA})\Big] \end{equation}

Now we are ready to derive differential equations for the systems evolution in time.

\subsection{Pair Approximation}

The game between real and fake news is a coordination game, and because of this, individuals will tend to form clusters of like-minded individuals, as observed in simulations. However, because of this, the probabilities along two successive edges are not independent. That is to say, if $p_{S_1S_2S_3}$ is the probability of starting at an $S_1$ player, following a random edge to an $S_2$ player, and then following another random edge to an $S_3$ player, we do \textbf{not} get that 
\begin{equation} \label{eq:pairapprox}
p_{S_1S_2S_3} = \frac{p_{S_1S_2}p_{S_2S_3}}{p_{S_2}}
\end{equation}
However, this makes studying the system untenable. Pair approximation alleviates this problem by making the simplifying assumptions that edges are independent and therefore Equation \eqref{eq:pairapprox} holds.

In the death-birth process, an individual is chosen to ``die'' and a neighbor is chosen to replicate and take the deceased individuals place. However, if the two individuals are playing the same strategy, nothing in the population will have changed. The only way the system changes is if an $A$ individual takes the place of a $B$ individual or vice versa, so we focus on the frequency of these two events to study the system.

We use the modified update step where only one individual is replaced per time step. This slows down the system's evolution by a factor of $\frac{1}{N}$, but it has very little effect on the behavior of the system, and it makes the system much easier to approach analytically. With a discrete time step $\Delta t = \frac{1}{N}$ so that one individual is replaced per time step, the differential equations for $p_A$ and $p_{AA}$ are:
\begin{equation}\label{eq:padot}
\dot{p_A} = \frac{1}{N}\frac{E(\Delta n_A)}{\Delta t} = E(\Delta n_A)
\end{equation}
\begin{equation}\label{eq:paadot}
\dot{p_{AA}} = \frac{2}{kN}\frac{E(\Delta n_{AA})}{\Delta t} = \frac{2}{k}E(\Delta n_{AA})
\end{equation}

We first focus on computing $E(\Delta n_A)$. Because only one individual updates at a time, $E(\Delta n_A) = P(\Delta n_A = 1) - P(\Delta n_A = -1)$. $n_A$ increases by one when a $B$ player is replaced by an $A$ player, and $n_A$ decreases by one when an $A$ player is replaced by a $B$ player. We now derive the probability of an $A$ player replacing a $B$ player. The probability of $B$ invading $A$ follows by symmetry.

The $B$ player that is being replaced has $k$ neighbors, each of which can be an $A$, $B$, or $C$ player. Specifically, the focal $B$ player has $k_B^A$ $A$ neighbors, $k_B^B$ $B$ neighbors, and $k_B^C$ $C$ neighbors with probability 
\begin{equation} \frac{k!}{k_B^A! k_B^B! k_B^C!}q_{A|B}^{k_B^A}q_{B|B}^{k_B^B}q_{C|B}^{k_B^C} \end{equation}
and there is always the restriction that $k_B^A + k_B^B + k_B^C = k$.

Each of these neighbors has $k-1$ neighbors (not including the focal $B$ player) that are also multinomially distributed. An $A$-playing neighbor will have ${k'}_A^A$ $A$ neighbors, ${k'}_A^B$ $B$ neighbors, and ${k'}_A^C$ $C$ neighbors with probability
\begin{equation} \frac{(k-1)!}{{k'}_A^A! {k'}_A^B! {k'}_A^C!} q_{A|A}^{{k'}_A^A} q_{B|A}^{{k'}_A^B} q_{C|A}^{{k'}_A^C} \end{equation}
Here we used pair approximation, because we ignore the higher-order terms that might arise knowing that the $A$ player already has a $B$ neighbor.

Likewise, the $B$ and $C$ players neighboring the focal $B$ player have neighbors whose strategies are multinomially distributed. To determine the strategy the focal $B$ player will choose to imitate, we need to know the payoffs of all of the neighbors. 

An $A$ neighbor of the focal $B$ player who has ${k'}_A^A$ $A$ neighbors, ${k'}_A^B$ $B$ neighbors (not including the focal $B$ player), and ${k'}_A^C$ $C$ neighbors has payoff
\begin{equation} \pi_A = {k'}_A^A a + ({k'}_A^B + 1) b + {k'}_A^C \alpha \end{equation} 
and fitness 
\begin{equation} \label{eq:fA} f_A({k'}_A^A, {k'}_A^B, {k'}_A^C) = e^{\beta \pi_A} \end{equation}

The same quantities for the $B$ and $C$ neighbors work the same way.
\begin{equation} \pi_B = {k'}_B^A c + ({k'}_B^B + 1) d + {k'}_B^C \gamma \end{equation}
\begin{equation} \label{eq:fB} f_B({k'}_B^A, {k'}_B^B, {k'}_B^C) = e^{\beta \pi_B} \end{equation}
\begin{equation} \pi_C = {k'}_C^A 0 + ({k'}_C^B + 1) 0 + {k'}_C^C 0 = 0 \end{equation}
\begin{equation} f_C({k'}_C^A, {k'}_C^B, {k'}_C^C) = e^{\beta \pi_C} = 1 \end{equation}

We are interested in the focal $B$ player being replaced by an $A$ player. Because individuals choose who to copy proportional to fitness, the probability of the $B$ player selecting one of its $A$ neighbors is
\begin{equation} \frac{k_B^A f_A}{k_B^A f_A + k_B^B f_B + k_B^C f_C} \end{equation}

All that remains is to sum over all possible configurations of the $B$ player's neighbors and their neighbors and multiply by $p_B$ (the probability that a $B$ player is selected to update) to get the final probability $W_{AB}$ that a $B$ player is replaced by an $A$ player:

\begin{equation} \label{eq:WABlong}
\begin{split}
W_{AB} = p_B & \cdot \sum_{k_B^A + k_B^B + k_B^C = k} \frac{k!}{k_B^A! k_B^B! k_B^C!}q_{A|B}^{k_B^A}q_{B|B}^{k_B^B}q_{C|B}^{k_B^C}\\
& \cdot \sum_{{k'}_A^A + {k'}_A^B + {k'}_A^C = k-1}\frac{(k-1)!}{{k'}_A^A! {k'}_A^B! {k'}_A^C!} q_{A|A}^{{k'}_A^A} q_{B|A}^{{k'}_A^B} q_{C|A}^{{k'}_A^C}\\
& \cdot \sum_{{k'}_B^A + {k'}_B^B + {k'}_B^C = k-1}\frac{(k-1)!}{{k'}_B^A! {k'}_B^B! {k'}_B^C!} q_{A|B}^{{k'}_B^A} q_{B|B}^{{k'}_B^B} q_{C|B}^{{k'}_B^C}\\
& \cdot \sum_{{k'}_C^A + {k'}_C^B + {k'}_C^C = k-1}\frac{(k-1)!}{{k'}_C^A!{k'}_C^B! {k'}_C^C!} q_{A|C}^{{k'}_C^A} q_{B|C}^{{k'}_C^B} q_{C|C}^{{k'}_C^C}\\
& \cdot \frac{k_B^A f_A({k'}_A^A, {k'}_A^B+1, {k'}_A^C)}{\splitfrac{k_B^A f_A({k'}_A^A, {k'}_A^B+1, {k'}_A^C) + k_B^B f_B({k'}_B^A, {k'}_B^B+1, {k'}_B^C)}{ + k_B^C f_C({k'}_C^A, {k'}_C^B+1, {k'}_C^C)}}
\end{split}
\end{equation}

(Though it is difficult to typeset within the margins, note that this is a nested sum and not the product of four separate sums.) Likewise, $W_{BA}$, the probability of $B$ invading $A$, is

\begin{equation}\label{eq:WBAlong}
\begin{split}
W_{BA} = p_A & \cdot \sum_{k_A^A + k_A^B + k_A^C = k} \frac{k!}{k_A^A! k_A^B! k_A^C!}q_{A|A}^{k_A^A}q_{B|A}^{k_A^B}q_{C|A}^{k_A^C}\\
& \cdot \sum_{{k'}_A^A + {k'}_A^B + {k'}_A^C = k-1}\frac{(k-1)!}{{k'}_A^A! {k'}_A^B! {k'}_A^C!} q_{A|A}^{{k'}_A^A} q_{B|A}^{{k'}_A^B} q_{C|A}^{{k'}_A^C}\\
& \cdot \sum_{{k'}_B^A + {k'}_B^B + {k'}_B^C = k-1}\frac{(k-1)!}{{k'}_B^A! {k'}_B^B! {k'}_B^C!} q_{A|B}^{{k'}_B^A} q_{B|B}^{{k'}_B^B} q_{C|B}^{{k'}_B^C}\\
& \cdot \sum_{{k'}_C^A + {k'}_C^B + {k'}_C^C = k-1}\frac{(k-1)!}{{k'}_C^A!{k'}_C^B! {k'}_C^C!} q_{A|C}^{{k'}_C^A} q_{B|C}^{{k'}_C^B} q_{C|C}^{{k'}_C^C}\\
& \cdot \frac{k_A^B f_B({k'}_B^A+1, {k'}_B^B, {k'}_B^C)}{\splitfrac{k_A^A f_A({k'}_A^A+1, {k'}_A^B, {k'}_A^C) + k_A^B f_B({k'}_B^A+1, {k'}_B^B, {k'}_B^C)}{ + k_A^C f_C({k'}_C^A+1, {k'}_C^B, {k'}_C^C)}}
\end{split}
\end{equation}

Furthermore, when $B$ is invaded by $A$ it increases the number of $A-A$ pairs by $k_B^A$, so we can define $\phi^A_{AB}$ to be the expected value for the change in $A-A$ edges due to a $B$ player being invaded by an $A$ player. (The subscript describes the direction of invasion and the superscript determines which pair it corresponds to, so $\phi^A_{AB}$ means an $A$ player is replacing a $B$ player, and this term tells us about the change in $A-A$ pairs.) Like in \eqref{eq:WABlong}, we have

\begin{equation} \label{eq:phiAABlong}
\begin{split}
\phi^A_{AB} = p_B & \cdot \sum_{k_B^A + k_B^B + k_B^C = k} k_B^A \frac{k!}{k_B^A! k_B^B! k_B^C!}q_{A|B}^{k_B^A}q_{B|B}^{k_B^B}q_{C|B}^{k_B^C}\\
& \cdot \sum_{{k'}_A^A + {k'}_A^B + {k'}_A^C = k-1}\frac{(k-1)!}{{k'}_A^A! {k'}_A^B! {k'}_A^C!} q_{A|A}^{{k'}_A^A} q_{B|A}^{{k'}_A^B} q_{C|A}^{{k'}_A^C}\\
& \cdot \sum_{{k'}_B^A + {k'}_B^B + {k'}_B^C = k-1}\frac{(k-1)!}{{k'}_B^A! {k'}_B^B! {k'}_B^C!} q_{A|B}^{{k'}_B^A} q_{B|B}^{{k'}_B^B} q_{C|B}^{{k'}_B^C}\\
& \cdot \sum_{{k'}_C^A + {k'}_C^B + {k'}_C^C = k-1}\frac{(k-1)!}{{k'}_C^A!{k'}_C^B! {k'}_C^C!} q_{A|C}^{{k'}_C^A} q_{B|C}^{{k'}_C^B} q_{C|C}^{{k'}_C^C}\\
& \cdot \frac{k_B^A f_A({k'}_A^A, {k'}_A^B+1, {k'}_A^C)}{\splitfrac{k_B^A f_A({k'}_A^A, {k'}_A^B+1, {k'}_A^C) + k_B^B f_B({k'}_B^A, {k'}_B^B+1, {k'}_B^C)}{ + k_B^C f_C({k'}_C^A, {k'}_C^B+1, {k'}_C^C)}}
\end{split}
\end{equation}

Note that \eqref{eq:phiAABlong} only differs from \eqref{eq:WABlong} in a single $k_B^A$ term in the first line, which is there because we are interested in the expected value of the change in $A-A$ edges, and there are $k_B^A$ new $A-A$ edges being formed. Similarly, we can write down: 

\begin{equation}\label{eq:phiABAlong}
\begin{split}
\phi^A_{BA} = p_A & \cdot \sum_{k_A^A + k_A^B + k_A^C = k} k_A^A \frac{k!}{k_A^A! k_A^B! k_A^C!}q_{A|A}^{k_A^A}q_{B|A}^{k_A^B}q_{C|A}^{k_A^C}\\
& \cdot \sum_{{k'}_A^A + {k'}_A^B + {k'}_A^C = k-1}\frac{(k-1)!}{{k'}_A^A! {k'}_A^B! {k'}_A^C!} q_{A|A}^{{k'}_A^A} q_{B|A}^{{k'}_A^B} q_{C|A}^{{k'}_A^C}\\
& \cdot \sum_{{k'}_B^A + {k'}_B^B + {k'}_B^C = k-1}\frac{(k-1)!}{{k'}_B^A! {k'}_B^B! {k'}_B^C!} q_{A|B}^{{k'}_B^A} q_{B|B}^{{k'}_B^B} q_{C|B}^{{k'}_B^C}\\
& \cdot \sum_{{k'}_C^A + {k'}_C^B + {k'}_C^C = k-1}\frac{(k-1)!}{{k'}_C^A!{k'}_C^B! {k'}_C^C!} q_{A|C}^{{k'}_C^A} q_{B|C}^{{k'}_C^B} q_{C|C}^{{k'}_C^C}\\
& \cdot \frac{k_A^B f_B({k'}_B^A+1, {k'}_B^B, {k'}_B^C)}{\splitfrac{k_A^A f_A({k'}_A^A+1, {k'}_A^B, {k'}_A^C) + k_A^B f_B({k'}_B^A+1, {k'}_B^B, {k'}_B^C)}{ + k_A^C f_C({k'}_C^A+1, {k'}_C^B, {k'}_C^C)}}
\end{split}
\end{equation}

\begin{equation}\label{eq:phiBABlong}
\begin{split}
\phi^B_{AB} = p_B & \cdot \sum_{k_B^A + k_B^B + k_B^C = k} k_B^B \frac{k!}{k_B^A! k_B^B! k_B^C!}q_{A|B}^{k_B^A}q_{B|B}^{k_B^B}q_{C|B}^{k_B^C}\\
& \cdot \sum_{{k'}_A^A + {k'}_A^B + {k'}_A^C = k-1}\frac{(k-1)!}{{k'}_A^A! {k'}_A^B! {k'}_A^C!} q_{A|A}^{{k'}_A^A} q_{B|A}^{{k'}_A^B} q_{C|A}^{{k'}_A^C}\\
& \cdot \sum_{{k'}_B^A + {k'}_B^B + {k'}_B^C = k-1}\frac{(k-1)!}{{k'}_B^A! {k'}_B^B! {k'}_B^C!} q_{A|B}^{{k'}_B^A} q_{B|B}^{{k'}_B^B} q_{C|B}^{{k'}_B^C}\\
& \cdot \sum_{{k'}_C^A + {k'}_C^B + {k'}_C^C = k-1}\frac{(k-1)!}{{k'}_C^A!{k'}_C^B! {k'}_C^C!} q_{A|C}^{{k'}_C^A} q_{B|C}^{{k'}_C^B} q_{C|C}^{{k'}_C^C}\\
& \cdot \frac{k_B^A f_A({k'}_A^A, {k'}_A^B+1, {k'}_A^C)}{\splitfrac{k_B^A f_A({k'}_A^A, {k'}_A^B+1, {k'}_A^C) + k_B^B f_B({k'}_B^A, {k'}_B^B+1, {k'}_B^C)}{ + k_B^C f_C({k'}_C^A, {k'}_C^B+1, {k'}_C^C)}}
\end{split}
\end{equation}

\begin{equation}\label{eq:phiBBAlong}
\begin{split}
\phi^B_{BA} = p_A & \cdot \sum_{k_A^A + k_A^B + k_A^C = k} k_A^B \frac{k!}{k_A^A! k_A^B! k_A^C!}q_{A|A}^{k_A^A}q_{B|A}^{k_A^B}q_{C|A}^{k_A^C}\\
& \cdot \sum_{{k'}_A^A + {k'}_A^B + {k'}_A^C = k-1}\frac{(k-1)!}{{k'}_A^A! {k'}_A^B! {k'}_A^C!} q_{A|A}^{{k'}_A^A} q_{B|A}^{{k'}_A^B} q_{C|A}^{{k'}_A^C}\\
& \cdot \sum_{{k'}_B^A + {k'}_B^B + {k'}_B^C = k-1}\frac{(k-1)!}{{k'}_B^A! {k'}_B^B! {k'}_B^C!} q_{A|B}^{{k'}_B^A} q_{B|B}^{{k'}_B^B} q_{C|B}^{{k'}_B^C}\\
& \cdot \sum_{{k'}_C^A + {k'}_C^B + {k'}_C^C = k-1}\frac{(k-1)!}{{k'}_C^A!{k'}_C^B! {k'}_C^C!} q_{A|C}^{{k'}_C^A} q_{B|C}^{{k'}_C^B} q_{C|C}^{{k'}_C^C}\\
& \cdot \frac{k_A^B f_B({k'}_B^A+1, {k'}_B^B, {k'}_B^C)}{\splitfrac{k_A^A f_A({k'}_A^A+1, {k'}_A^B, {k'}_A^C) + k_A^B f_B({k'}_B^A+1, {k'}_B^B, {k'}_B^C)}{ + k_A^C f_C({k'}_C^A+1, {k'}_C^B, {k'}_C^C)}}
\end{split}
\end{equation}

Once we have these quantities (Equations \eqref{eq:WABlong} - \eqref{eq:phiBBAlong}), we have expressions for all of our independent variables.
\begin{equation} 
\dot{p_{CC}} = 0 
\end{equation}
\begin{equation} \label{eq:papbdot}
\dot{p_A} = -\dot{p_B}= W_{AB} - W_{BA} 
\end{equation}
\begin{equation} \label{eq:paadotshort}
\dot{p_{AA}} = \frac{2}{k}(\phi^A_{AB} - \phi^A_{BA}) 
\end{equation}
\begin{equation} \label{eq:pbbdotshort}
\dot{p_{BB}} = \frac{2}{k}(\phi^B_{BA} - \phi^B_{AB}) 
\end{equation}

\subsection{Weak Selection}

Even with the substantial simplification from pair approximation, the previous results are too complicated and unwieldy to be useful by themselves. Because of  compounding sums, directly calculating the derivatives requires adding millions of terms if $k=8$. Furthermore, the pair approximation means that we lose the information critical to clustering, and therefore the analytic results here will fail to capture the pseudo-steady states that we observe when $\beta$ is much larger than zero.

We can sidestep both these issues by working in the limit of weak selection. In weak selection, the success or failure of an individual in the fake news game is only one small factor in the individual's success, and fitnesses are much more uniform across the population. When $\beta$ is close to zero, we can throw out higher order terms which simplifies the expression, and when $\beta$ is close to zero, the pseudo-steady states cannot exist anyways because the system behaves approximately like neutral drift. Taking the Taylor expansion of the exponential in equations \eqref{eq:fA} and \eqref{eq:fB} with respect to $\beta$ and only keeping the low order terms, what is left is mathematically tractable. We have expressions for each of $W_{AB}, W_{BA}, \phi_{AB}^A, \phi_{BA}^A, \phi_{BA}^B, \phi_{AB}^B$. We manipulate each separately and bring them back together at the end.

\subsection{$W_{AB}$ and $W_{BA}$:}
Equation \eqref{eq:WABlong} gives us an expression for $W_{AB}$. The fact-checkers playing $C$ have constant fitness, $f_C = 1$, and no other terms in the last line of \eqref{eq:WABlong} depend on the neighbors of $C$ players, so we can pull it all through the final sum which collapses to 1 because it is the sum of the probabilities of all possible configurations of neighbors, which must be 1. Therefore,
\begin{equation}\label{WABshort}
\begin{split}
W_{AB} = p_B & \cdot \sum_{k_B^A + k_B^B + k_B^C = k} \frac{k!}{k_B^A! k_B^B! k_B^C!} q_{A|B}^{k_B^A} q_{B|B}^{k_B^B} q_{C|B}^{k_B^C} \\
&\cdot \sum_{k_A^{A'} + k_A^{B'} + k_A^{C'} = k-1}\frac{(k-1)!}{k_A^{A'}! k_A^{B'}! k_A^{C'}!} q_{A|A}^{k_A^{A'}} q_{B|A}^{k_A^{B'}} q_{C|A}^{k_A^{C'}}\\
&\cdot \sum_{k_B^{A'} + k_B^{B'} + k_B^{C'} = k-1}\frac{(k-1)!}{k_B^{A'}! k_B^{B'}! k_B^{C'}!} q_{A|B}^{k_B^{A'}} q_{B|B}^{k_B^{B'}} q_{C|B}^{k_B^{C'}}\\
& \cdot \frac{k_B^A f_A(k_A^{A'}, k_A^{B'}+1, k_A^{C'})}{k_B^A f_A(k_A^{A'}, k_A^{B'}+1, k_A^{C'}) + k_B^B f_B(k_B^{A'}, k_B^{B'}+1, k_B^{C'}) + k_B^C} 
\end{split}
\end{equation}
Then, using the Taylor expansion for the exponentials in $f_A$ and $f_B$ but only keeping the low order terms of $\beta$, we have
\begin{equation} \label{eq:WABloworder}
\begin{split}
&\frac{k_B^A f_A(k_A^{A'}, k_A^{B'}+1, k_A^{C'})}{k_B^A f_A(k_A^{A'}, k_A^{B'}+1, k_A^{C'}) + k_B^B f_B(k_B^{A'}, k_B^{B'}+1, k_B^{C'}) + k_B^C} \\
&\approx \frac{k_B^A \big(1+\beta(a k_A^{A'} + b (k_A^{B'} + 1) + c k_A^{C'})\big)}{\splitfrac{k_B^A \big(1+\beta(a k_A^{A'} + b (k_A^{B'} + 1) + \alpha k_A^{C'})\big)}{ + k_B^B\big(1 + \beta(c k_B^{A'} + d (k_B^{B'} + 1) + \gamma k_B^{C'})\big) + k_B^C}} \\
&= \frac{k_B^A(1+\beta u_1)}{k + \beta(k_B^A u_1 + k_B^B u_2)}\\
&\approx k_B^A(1+\beta u_1)[\frac{1}{k} - \frac{k_B^A u_1 + k_B^B u_2}{k^2} \beta] \\
&\approx \frac{k_B^A}{k} + \beta[\frac{k_B^A u_1}{k} - k_B^A \frac{k_B^A u_1 + k_B^B u_2}{k^2}]
\end{split}
\end{equation}

where $u_1 = a k_A^{A'} + b (k_A^{B'} + 1) + \alpha k_A^{C'}$ and $u_2 = c k_B^{A'} + d (k_B^{B'} + 1) + \gamma k_B^{C'}$.
By carefully pulling terms through the sums, we have the following identities:

\begin{equation}
\begin{split}
\sum_{k_A^{A'} + k_A^{B'} + k_A^{C'} = k-1}&\frac{(k-1)!}{k_A^{A'}! k_A^{B'}! k_A^{C'}!} q_{A|A}^{k_A^{A'}} q_{B|A}^{k_A^{B'}} q_{C|A}^{k_A^{C'}} u_1 \\
&= a(k-1) q_{A|A} + b\big((k-1)q_{B|A}+1\big) + \alpha(k-1) q_{C|A} \\
&= E_A+b 
\end{split}
\end{equation}

\begin{equation} 
\begin{split}
\sum_{k_B^{A'} + k_B^{B'} + k_B^{C'} = k-1}&\frac{(k-1)!}{k_B^{A'}! k_B^{B'}! k_B^{C'}!} q_{A|B}^{k_B^{A'}} q_{B|B}^{k_B^{B'}} q_{C|B}^{k_B^{C'}} u_2 \\
&= c(k-1)q_{A|B} + d\big((k-1)q_{B|B}+1\big) +\gamma (k-1) q_{C|B} \\
&= E_B+d
\end{split}
\end{equation}

Notice that $E_A$ and $E_B$ are the expected payoffs for $A$ and $B$ players from $k-1$ neighbors. Using these identities on our equation for $W_{AB}$, we get that
\begin{equation}
\begin{split}
W_{AB} = p_B& \cdot \sum_{k_B^A + k_B^B + k_B^C = k} \frac{k!}{k_B^A! k_B^B! k_B^C!} q_{A|B}^{k_B^A} q_{B|B}^{k_B^B} q_{C|B}^{k_B^C} \\
&\cdot \Bigg[ \frac{k_B^A}{k} - \beta\frac{k_B^Bk_B^A}{k^2}(E_B+d) + \beta \frac{k_B^A}{k} (E_A+b) - \beta \frac{{k_B^A}^2}{k^2} (E_A+b) \Bigg]
\end{split}
\end{equation}

Each of these four terms in the brackets can be dealt with separately in similar fashion:
\begin{equation}\label{eq:sum1}
\sum_{k_B^A + k_B^B + k_B^C = k} \frac{k!}{k_B^A! k_B^B! k_B^C!} q_{A|B}^{k_B^A} q_{B|B}^{k_B^B} q_{C|B}^{k_B^C} \Bigg[\frac{k_B^A}{k}\Bigg] = q_{A|B}
\end{equation}
\begin{equation}
\begin{split}
\sum_{k_B^A + k_B^B + k_B^C = k} &\frac{k!}{k_B^A! k_B^B! k_B^C!} q_{A|B}^{k_B^A} q_{B|B}^{k_B^B} q_{C|B}^{k_B^C}\Bigg[ - \beta\frac{k_B^Bk_B^A}{k^2}(E_B+d) \Bigg] \\ &= - \beta\frac{(E_B+d)}{k^2}k(k-1)q_{A|B} q_{B|B}
\end{split}
\end{equation}
\begin{equation}
\sum_{k_B^A + k_B^B + k_B^C = k} \frac{k!}{k_B^A! k_B^B! k_B^C!} q_{A|B}^{k_B^A} q_{B|B}^{k_B^B} q_{C|B}^{k_B^C} \Bigg[ \beta \frac{k_B^A}{k} (E_A+b)\Bigg] = \beta (E_A+b) q_{A|B}
\end{equation}
\begin{equation}\label{eq:sum4}
\begin{split}
\sum_{k_B^A + k_B^B + k_B^C = k} &\frac{k!}{k_B^A! k_B^B! k_B^C!} q_{A|B}^{k_B^A} q_{B|B}^{k_B^B} q_{C|B}^{k_B^C} \Bigg[ - \beta \frac{{k_B^A}^2}{k^2} (E_A+b) \Bigg] \\ &= -\beta\frac{(E_A+b)}{k^2} k q_{A|B}[(k-1)q_{A|B} + 1] 
\end{split}
\end{equation}

Therefore,
\begin{equation}
\begin{split}
W_{AB} = p_B&\bigg[ q_{A|B} + \beta \Big( (E_A + b) q_{A|B} - \frac{E_A + b}{k} q_{A|B} \\ &- \frac{k-1}{k} q_{A|B}\big[ (E_B + d)q_{B|B} + (E_A+b) q_{A|B} \big] \Big) \bigg] + \mathcal{O}(\beta^2)
\end{split}
\end{equation}

Using the same techniques, we can simplify our expression for $W_{BA}$:

\begin{equation} 
\begin{split}
W_{BA} = p_A&\bigg[ q_{B|A} + \beta \Big( (E_B + c) q_{B|A} - \frac{E_B + c}{k} q_{B|A} \\& - \frac{k-1}{k} q_{B|A}\big[ (E_B + c)q_{B|A} + (E_A+a) q_{A|A} \big] \Big) \bigg] + \mathcal{O}(\beta^2)
\end{split}
\end{equation}

Note immediately that since $p_B q_{A|B} = p_A q_{B|A}$, the zero-th order terms of $W_{AB}$ and $W_{BA}$ are equal.

\subsection{The $\phi$s:}
The pair derivatives are non-zero, even when $\beta = 0$, so we will focus only on the zeroth order terms, because these will dominate the first-order terms when $\beta$ is small. 
\begin{equation}
\begin{split}
\phi_{AB}^A = p_B \cdot &\sum_{k_B^A + k_B^B + k_B^C = k} k_B^A\frac{k!}{k_B^A! k_B^B! k_B^C!} q_{A|B}^{k_B^A} q_{B|B}^{k_B^B} q_{C|B}^{k_B^C} \\
&\cdot \sum_{k_A^{A'} + k_A^{B'} + k_A^{C'} = k-1}\frac{(k-1)!}{k_A^{A'}! k_A^{B'}! k_A^{C'}!} q_{A|A}^{k_A^{A'}} q_{B|A}^{k_A^{B'}} q_{C|A}^{k_A^{C'}}\\
&\cdot \sum_{k_B^{A'} + k_B^{B'} + k_B^{C'} = k-1}\frac{(k-1)!}{k_B^{A'}! k_B^{B'}! k_B^{C'}!} q_{A|B}^{k_B^{A'}} q_{B|B}^{k_B^{B'}} q_{C|B}^{k_B^{C'}}\\
&\cdot \sum_{k_C^{A'} + k_C^{B'} + k_C^{C'} = k-1}\frac{(k-1)!}{k_C^{A'}! k_C^{B'}! k_C^{C'}!} q_{A|C}^{k_C^{A'}} q_{B|C}^{k_C^{B'}} q_{C|C}^{k_C^{C'}}\\
& \cdot \frac{k_B^A f_A(k_A^{A'}, k_A^{B'}+1, k_A^{C'})}{\splitfrac{k_B^A f_A(k_A^{A'}, k_A^{B'}+1, k_A^{C'}) + k_B^B f_B(k_B^{A'}, k_B^{B'}+1, k_B^{C'})}{ + k_B^C f_C(k_C^{A'}, k_C^{B'}+1, k_C^{C'})}}
\end{split}
\end{equation}
The zeroth order terms are what is left when $\beta = 0$, or when we have neutral drift. In that case, $f_A = f_B = f_C= 1$, and most of the sums collapse to 1. We quickly get that
\begin{equation}
\phi_{AB}^A = \frac{p_B}{k} \sum_{k_B^A + k_B^B + k_B^C = k} \frac{k!}{k_B^A! k_B^B! k_B^C!} q_{A|B}^{k_B^A} q_{B|B}^{k_B^B} q_{C|B}^{k_B^C}{k_B^A}^2
\end{equation}
We relabel  for notational convenience and readability when evaluating this sum. Let $X = k_B^A, Y = k_B^B, Z = k_B^C$. Then the sum is 
\begin{equation}
\begin{split}
&\sum_{X+Y+Z = k} \frac{k!}{X!Y!Z!} q_{A|B}^X q_{B|B}^Y q_{C|B}^Z X^2 \\ = & k q_{A|B}\sum_{(X-1)+Y+Z = k-1} \frac{(k-1)!}{(X-1)!Y!Z!}	q_{A|B}^{X-1} q_{B|B}^Y q_{C|B}^Z (X) \\
=&   k q_{A|B}\sum_{(X-1)+Y+Z = k-1} \frac{(k-1)!}{(X-1)!Y!Z!}	q_{A|B}^{X-1} q_{B|B}^Y q_{C|B}^Z (X-1) \\
&+  k q_{A|B}\sum_{(X-1)+Y+Z = k-1} \frac{(k-1)!}{(X-1)!Y!Z!}	q_{A|B}^{X-1} q_{B|B}^Y q_{C|B}^Z \\
=& k q_{A|B} \Big( (k-1)q_{A|B} \sum_{(X-2)+Y+Z = k-2} \frac{(k-2)!}{(X-2)!Y!Z!}	q_{A|B}^{X-2} q_{B|B}^Y q_{C|B}^Z +1 \Big) \\
=& k q_{A|B} \Big( (k-1)q_{A|B} + 1\Big)
\end{split}
\end{equation}

Immediately, we get,
\begin{equation}\label{eq:phiABAshort}
\phi_{AB}^A = p_B q_{A|B} \Big( (k-1) q_{A|B} + 1\Big) + \mathcal{O}(\beta)
\end{equation}

The other $\phi$ terms are calculated in the same way. They are:
\begin{equation}\label{eq:phiBAAshort}
\phi_{BA}^A = p_A (k-1) q_{A|A} q_{B|A} + \mathcal{O}(\beta)
\end{equation}
\begin{equation}\label{eq:phiBABshort}
\phi_{BA}^B = p_A q_{B|A} \Big( (k-1) q_{B|A} + 1 \Big) + \mathcal{O}(\beta)
\end{equation}
\begin{equation}\label{eq:phiABBshort}
\phi_{AB}^B = p_B (k-1) q_{B|B} q_{A|B} + \mathcal{O}(\beta)
\end{equation}

\subsection{The Slow Manifold}

With these simplified equations, we can solve the system. Consider the zero-th order terms, setting $\beta = 0$. $W_{AB} = W_{BA}$, so $\dot{p}_A = \dot{p} _B = \dot{p_C}$. Now we address $\dot{p_{AA}}$, $\dot{p_{BB}}$, and $\dot{p_{AB}}$:

With the above derivatives and \eqref{eq:pab}, we get that 
\begin{equation} \label{eq:pabdot}
\dot{p_{AB}} =-\frac{1}{2}( \dot{p_{AA}} + \dot{p_{AA}})
\end{equation}

By substituting \eqref{eq:phiABAshort} and \eqref{eq:phiBAAshort} into \eqref{eq:paadotshort}:
\begin{equation} \label{eq:paadotsimp}
\begin{split}
\dot{p_{AA}} &= \frac{2}{k}\Big[\phi^A_{AB} - \phi^A_{BA}\Big]\\
&= \frac{2}{k} \Big[ p_B q_{A|B} \Big( (k-1) q_{A|B} + 1\Big) - p_A (k-1) q_{A|A} q_{B|A} \Big]\\
&= \frac{2}{k} \Big[ p_B q_{A|B}q_{A|B}(k-1) - p_A q_{A|A}q_{B|A}(k-1) + p_B q_{B|A} \Big]\\
&=  \frac{2}{k} \Big[ \frac{p_{AB}^2}{p_B}(k-1) - \frac{p_{AA}p_{AB}}{p_A}(k-1) + p_{AB} \Big]
\end{split}
\end{equation}
Similarly, with \eqref{eq:phiBABshort} and \eqref{eq:phiABBshort} in \eqref{eq:pbbdotshort}:
\begin{equation}\label{eq:pbbdotsimp}
\begin{split}
\dot{p_{BB}} &= \frac{2}{k} \Big[\phi^B_{BA} - \phi^B_{AB}\Big]\\
&= \frac{2}{k} \Big[p_Aq_{B|A}\Big( (k-1)q_{B|A}+1 \Big) - p_B(k-1)q_{B|B}q_{A|B}\Big]\\
&=\frac{2}{k} \Big[ p_A q_{B|A}q_{B|A}(k-1) - p_B q_{B|B}q_{A|B}(k-1) + p_Aq_{B|A}\Big]\\
&= \frac{2}{k} \Big[ \frac{p_{AB}^2}{p_A}(k-1) - \frac{p_{AB}p_{BB}}{p_B}(k-1) + p_{AB} \Big]
\end{split}
\end{equation}
Now subtract \eqref{eq:pbbdotsimp} from \eqref{eq:paadotsimp}:
\begin{equation}\label{eq:difdot}
\begin{split}
\dot{p_{AA}}-\dot{p_{BB}} =& \frac{2}{k}\Big[ \frac{p_{AB}^2}{p_B}(k-1) - \frac{p_{AA}p_{AB}}{p_A}(k-1) + p_{AB}\Big]\\
& - \frac{2}{k}\Big[ \frac{p_{AB}^2}{p_A}(k-1) - \frac{p_{AB}p_{BB}}{p_B}(k-1) + p_{AB} \Big]\\
=& \frac{2(k-1)}{k} \Big[ \frac{p_{AB}^2}{p_B} - \frac{p_{AA}p_{AB}}{p_A} - \frac{p_{AB}^2}{p_A} + \frac{p_{AB}p_{BB}}{p_B} \Big]
\end{split}
\end{equation}

When the system is initialized at $t=0$, it is well-mixed and $p_{S_1S_2}(0) = p_{S_1}(0)p_{S_2}(0)$ for all strategies $S_1$ and $S_2$. Thus, at $t=0$, by equation \eqref{eq:difdot}, $\dot{p_{AA}}-\dot{p_{BB}}=0$. And together with \eqref{eq:pabdot}, we have
\begin{equation}\label{eq:eqdot}
\dot{p_{AA}}=\dot{p_{BB}}=-\dot{p_{AB}}
\end{equation}

In fact, this will hold for all time steps, because as long as it holds, it will continue to hold. A sketch of a formal proos is as follows: solve the system with Euler's method and take the limit as the discrete time step goes to zero. By the convergence of Euler's method, \eqref{eq:eqdot} holds for all $t$.

From this, \eqref{eq:pac} and \eqref{eq:pbc} show that $\dot{p_{AC}} = \dot{p_{BC}} = 0$. Then, 
\begin{equation}
\dot{q_{C|A}} = \frac{d}{dt}\frac{p_{AC}}{p_A} = \frac{\dot{p_{AC}}p_A-p_{AC}\dot{p_A}}{p_A^2} = 0
\end{equation}
Similarly, $\dot{q_{C|B}}=0$. These results are expected because in neutral drift, the fact-checkers do not give either strategy an advantage, so fact-checkers will not naturally attract $A$ players or repel $B$ players.

Because $\beta$ is very small, the zero-th order terms in $\dot{p}_{AA}$ and $\dot{p}_{BB}$ will go to zero much quicker than the first order terms in $\dot{p}_A$ and $\dot{p}_B$. Set $\dot{p}_{AA} = 0$:
\begin{equation} 
\dot{p}_{AA} = \frac{2}{k} \Big[p_B q_{A|B} \Big( (k-1)q_{A|B} + 1 \Big) - p_A (k-1) q_{A|A} q_{B|A}\Big] = 0
\end{equation}
Rearranging and dividing by $\frac{2 p_{AB}}{k}$ gives
\begin{equation}
(k-1) q_{A|B} + 1 = (k-1) q_{A|A}
\end{equation}
Now use the identities $q_{A|B} = \frac{p_A}{p_B} q_{B|A}$ and $q_{B|A} = 1 - q_{A|A} - p_C$ and rearrange to get
\begin{equation}\label{eq:qaa}
q_{A|A} = p_A + \frac{p_B}{(k-1)(1-p_C)}
\end{equation}
A similar procedure with $\dot{p}_{BB}=0$ yields
\begin{equation}\label{eq:qbb}
q_{B|B} = p_B + \frac{p_A}{(k-1)(1-p_C)}
\end{equation}

These conditions define the slow manifold, where the system changes slowly due to $\beta$ being close to zero. The system may start as a well-mixed population, but it will very quickly approach a state where the above conditions hold, at least approximately. Notice that the slow manifold is one-dimensional; everything can be expressed in terms of $p_A$, because $p_B = 1 - p_C - p_A$, and $p_C$ will be a constant.

\subsection{Fixation Probabilities}

Consider a system starting with $p_A(0) = p$ and a small time step $\Delta t$ in which we assume one death-birth occurs. Renormalize with ${p_A}_{\text{new}} = {p_A}_{\text{old}}/(1-p_C)$ and ${p_B}_{\text{new}} = {p_B}_{\text{old}}/(1-p_C)$ so that $p_A$ and $p_B$ are between 0 and 1. Now $p_A$ and $p_B$ represent the proportion of individuals playing $A$ or $B$ out of all the individuals that are capable of changing their strategy (the $A$ and $B$ players). There is a mean $m_A(p)$ and variance $v_A(p)$ of $\Delta p_A$ for a single time step. We have 
\begin{equation}\label{eq:map}
m_A(p) = E(\Delta p_A) = \frac{1}{N} [W_{AB} - W_{BA}] = [W_{AB} - W_{BA}]\Delta t
\end{equation}
\begin{equation}\label{eq:vap}
\begin{split}
v_A(p) &= E(\Delta p_A^2) - E(\Delta p_A)^2 = E(\Delta p_A^2) + \mathcal{O}(\beta^2)\\ &\approx \frac{1}{N^2} [W_{AB}+W_{BA}] = \frac{1}{N} [W_{AB}+W_{BA}]\Delta t
\end{split}
\end{equation}

The relevant value will be $-\frac{2m_A(p)}{v_A(p)}$, which can be obtained by substituting in the constraints of the slow manifold: Equations \eqref{eq:qaa} and \eqref{eq:qbb}. After substituting in the expressions for $W_{AB}$ and $W_{BA}$, simplifying gets us:
\begin{equation}
-\frac{2m_A(p)}{v_A(p)} = \frac{\beta N}{k}(u_1 p + u_2)
\end{equation}

where

\begin{equation}
u_1 = (a-b-c+d)(1-k^2 - \frac{1+k}{p_C-1})(1-p_C)
\end{equation}

\begin{equation}\label{eq:u2}
u_2 = -a+b+c-d-ak+bk-bk^2+dk^2+(k-1)\big(c+(b-\alpha+\gamma)k-d(1+k)\big)p_C
\end{equation}

According to diffusion theory, the fixation probability of $A$ beginning with $p_A(0)=p$, denoted $\rho_A(p)$, satisfies the equation
\begin{equation}
m_A(p) \frac{d \rho_A(p)}{dp} + \frac{v_A(p)}{2} \frac{d^2 \rho_A(p)}{dp^2} = 0
\end{equation}
This equation is separable and first order with respect to $\frac{d \rho_A(p)}{dp}$. 
\begin{equation}
\ln \frac{d \rho_A(p)}{dp} = \int - \frac{2m_A(p)}{v_A(p)} dp
\end{equation}
The low order terms are 
\begin{equation}
\frac{d \rho_A(p)}{dp} = 1 + \frac{\beta N}{k}\Big(\frac{u_1}{2} p^2 + u_2 p \Big) + c_1
\end{equation}
$c_1$ is a constant of integration.
Integrating once more gives
\begin{equation}
\rho_A(p) = p +  \frac{\beta N}{k}\Big(\frac{u_1}{6} p^3 + \frac{u_2}{2} p^2\Big) + c_1 p + c_2
\end{equation}
Using the boundary conditions $\rho_A(0) = 0$ and $\rho_A(1) = 1$ to solve for the constants of integration, we get 
\begin{equation}
\begin{split}\label{eq:pap}
\rho_A(p) =& p + \frac{\beta N}{k} \Big( \frac{u_1}{6}p^3 + \frac{u_2}{2} p^2 - (\frac{u_1}{6} + \frac{u_2}{2})p \Big)\\
=& p + \frac{\beta N p (1-p)}{6k}\Big(-3u_2 - u_1(1+p)\Big)
\end{split}
\end{equation}
When $p\ll1$, such as when $p=1/N$ for invasion probabilities, \eqref{eq:pap} becomes
\begin{equation} \label{eq:papapprox}
\rho_A(p) \approx p + \frac{\beta N p (1-p)}{6k}\Big(-3u_2 - u_1\Big)
\end{equation}

We can use this work to calculate the fixation probability for the $B$ strategy, as well. For a given $p_A$ and $p_B$ with $p_A + p_B = 1$, $m_B(p) = - m_A(1-p)$ and $v_B(p) = v_A(1-p)$. Therefore
\begin{equation}
\frac{-2 m_B(p)}{v_B(p)} = \frac{2 m_A(1-p)}{v_A(1-p)} = - \frac{\beta N}{k}\Big(u_1 ( 1 - p) + u_2\Big) = \frac{\beta N}{k}\Big(u_1 p - (u_1 + u_2)\Big)
\end{equation}

From this, as in \eqref{eq:papapprox},
\begin{equation}\label{eq:pbpapprox}
\rho_B(p) = p + \frac{\beta N}{k} \Big( \frac{w_1}{6}p^3 + \frac{w_2}{2}p^2 - (\frac{w_1}{6} + \frac{w_2}{2})p \Big)\approx p + \frac{\beta N}{k} \Big(-w_1 - 3 w_2\Big)
\end{equation}
with $w_1 = u_1$ and $w_2 = -(u_2 + u_1)$.

\subsection{Fact-checker Accuracy}

The adjustment to include a parameter $\lambda$ to take into account inaccurate fact-checkers is very simple. Recall that a fact-checker with accuracy $\lambda \in [0,1]$ gives benefit $\alpha$ to an $A$ player and penalty $\gamma$ to a $B$ player with probability $\lambda$, and gives the opposite payoffs with probability $1 - \lambda$. Therefore, the expected payoff an $A$ player receives from a $C$ player is $\lambda \alpha + (1-\lambda)\gamma$ and the expected payoff for a $B$ player is $\lambda \gamma + (1-\lambda) \alpha$. 

All the previous work with pair approximation, weak selection, and the diffusion approximation still hold, but we can replace the old expected payoffs of $\alpha$ and $\gamma$ with the new expected payoffs $\lambda \alpha + (1-\lambda)\gamma$ and $\lambda \gamma + (1-\lambda) \alpha$, respectively. Conveniently, the substitution can be done at the very end, where $\alpha$ and $\gamma$ appear as coefficients in Equation \eqref{eq:u2}.